\documentclass[onecollarge,smallextended,numbook]{svjour2}
\def\makeheadbox{}
   
\usepackage{color}
\usepackage[hidelinks]{hyperref}
\usepackage{graphicx}
\usepackage{amsfonts}

\def\Eq#1{\label{#1}}
\def\equ#1{(\ref{#1})}

\definecolor{iblue}{RGB}{65,105,225}
\definecolor{ired}{RGB}{220,20,60}
\definecolor{igreen}{RGB}{50,205,50}
\definecolor{ipurple}{RGB}{75,0,130}
\definecolor{iochre}{RGB}{218,165,32}
\definecolor{iteal}{RGB}{51,204,204}
\definecolor{imauve}{RGB}{204,51,153}

\let\a=\alpha \let\b=\beta        \let\d=\delta  
\let\z=\zeta  \let\h=\eta     \let\k=\kappa  \let\l=\lambda
\let\m=\mu    \let\n=\nu     \let\x=\xi        \let\p=\pi     
\let\s=\sigma \let\t=\tau    \let\f=\varphi    \let\ch=\chi
\let\ch=\chi  \let\ps=\psi      \let\o=\omega  
 \let\D=\Delta

\def\V#1{{\bf#1}}
\def\*{\vskip 3mm}
\def\0{\noindent}
\def\be{\begin{equation}}
\def\ee{\end{equation}}
\def\bea{\begin{eqnarray}}
\def\eea{\end{eqnarray}}
\def\nn{\nonumber}

\def\HH{{\cal H}}
\def\NN{{\cal N}}
\def\RR{{\cal R}}
\let\dpr=\partial\let\fra=\frac
\def\ie{{\it i.e.}\ }
\def\eg{{\it e.g.}\ }

\def\defi{{\buildrel def\over=}}

\def\media#1{{\Blangle\,#1\,\Brangle}}
\def\ket#1{{\Bvert#1\Brangle}}

\def\otto{\,{\kern-1.truept\leftarrow\kern-5.truept\to\kern-1.truept}\,}
\def\wt#1{\widetilde{#1}}

\def\tende#1{\,\vtop{\ialign{##\crcr\rightarrowfill\crcr
 \noalign{\kern-1pt\nointerlineskip} \hskip3.pt${\scriptstyle
 #1}$\hskip3.pt\crcr}}\,}
\def\ie{{\it i.e.\ }}

\def\Ba   {{\mbox{\boldmath$ \alpha$}}}

\def\Bs   {{\mbox{\boldmath$ \sigma$}}}

\def\Bt   {{\mbox{\boldmath$ \tau$}}}

\def\Bo   {{\mbox{\boldmath$ \omega$}}}

\def\Brangle {{\mbox{\boldmath$ \rangle$}}}
\def\Blangle {{\mbox{\boldmath$ \langle$}}}
\def\Bvert{{\mbox{\boldmath$|$}}}
\def\Bell   {{\mbox{\boldmath$\ell$}}}
\let\up\uparrow
\let\down\downarrow
\def\({\left(}
\def\){\right)}

\begin{document}

\title{Kondo effect in a fermionic hierarchical model}

\author{Giuseppe Benfatto \and Giovanni Gallavotti \and Ian Jauslin}

\institute{
Giuseppe Benfatto \at 
Universit\`a degli studi di Roma ``Tor Vergata'',\\ 
Via della Ricerca Scientifica 1,\\ 00133 Roma, Italy\\
\email{benfatto@mat.uniroma2.it}\\
homepage {http://axp.mat.uniroma2.it/$\sim$benfatto}
\and
Giovanni Gallavotti \at
INFN-Roma1 and Rutgers University,\\
P.le Aldo Moro 2,\\
00185 Roma, Italy\\
\email{giovanni.gallavotti@roma1.infn.it}\\
homepage {http://ipparco.roma1.infn.it/$\sim$giovanni}
\and
Ian Jauslin \at
University of Rome ``La Sapienza'', Dipartimento di Fisica,\\
P.le Aldo Moro 2,\\
00185 Roma, Italy\\
\email{ian.jauslin@roma1.infn.it}\\
homepage {http://ian.jauslin.org/}
}

\date{June 22, 2015}

\maketitle

\begin{abstract}
In this paper, a {\it fermionic hierarchical model}
is defined, inspired by the {\it Kondo model}, which describes a
1-dimensional lattice gas of spin-1/2 electrons interacting with a
spin-1/2 impurity.  This model is proved to be {\it exactly solvable},
and is shown to exhibit a {\it Kondo effect}, i.e. that, if the
interaction between the impurity and the electrons is antiferromagnetic,
then the magnetic susceptibility of the impurity is finite in the
0-temperature limit, whereas it diverges if the interaction is
ferromagnetic. Such an effect is therefore inherently non-perturbative.
This difficulty is overcome by using the exact solvability of the model,
which follows both from its fermionic and hierarchical nature.
\end{abstract}

\keywords{Renormalization group\and Non-perturbative renormalization\and
  Kondo effect\and Fermionic hierarchical model\and Quantum field theory}

\section{Introduction}
\label{sec1}
%%%%%%%%%%%%%%%%%%%%%%%%%%%

Although at high temperature the resistivity of most metals is an
increasing function of the temperature, experiments carried out since the
early XX\textsuperscript{th} century have shown that in metals containing
trace amounts of magnetic impurities (i.e. copper polluted by iron), the
resistivity has a minimum at a small but positive temperature, below which
the resistivity decreases as the temperature increases. One interesting
aspect of such a phenomenon, is its strong non-perturbative nature: it has
been measured in samples of copper with iron impurities at a concentration
as small as 0.0005\% \cite{Ko005}, which raises the question of how such a
minute perturbation can produce such an effect.  Kondo introduced a toy
model in 1964, see Eq.\equ{e2.1} below, to understand such a phenomenon,
and computed electronic scattering amplitudes at {third} order in the Born
approximation scheme \cite{Ko964}, and found that the effect may stem
from an antiferromagnetic coupling between
the impurities (called ``localized spins'' in \cite{Ko964})
and the electrons in the metal. The existence of such a coupling had been
proposed by Anderson \cite{An961}.

Kondo's theory attracted great attention and its scaling properties and
connection to $1D$ Coulomb gases were understood
\cite{Dy969,An970,AYH970b}
\footnote{\small The obstacle to a complete understanding of the model
  (with $\l_0<0$) being what would later be called the growth of a relevant
  coupling.}
when in a seminal paper, published in 1975 \cite{Wi975},
Wilson addressed and solved the problem by constructing a sequence of Hamiltonians that
adequately represent the system on ever increasing length scales. Using
ideas from his formulation of the renormalization group, Wilson showed,
by a combination of numerical and perturbative methods,  that only few
(three) terms in each Hamiltonian, need to be studied in order to
account for the Kondo effect (or rather, a related effect on the magnetic
susceptibility of the impurities, see below).

The non-perturbative nature of the effect manifests itself in Wilson's
formalism by the presence of a non-trivial fixed point in the renormalization
group flow, at which the corresponding effective theory behaves in a way that
is qualitatively different from the non-interacting one. Wilson has studied the
system around the non-trivial fixed point by perturbative expansions, but the
intermediate regime (in which perturbation theory breaks down)
was studied by numerical methods. In fact, when using renormalization group
techniques to study systems with non-trivial fixed points, oftentimes one cannot
treat non-perturbative regimes analytically. The hierarchical Kondo model, which
will be discussed below, is an exception to this rule: indeed, we will show
that the physical properties of the model can be obtained by iterating an
{\it explicit} map, computed analytically, and called the {\it beta function},
whereas, in the current state of the art, the beta function for the full
(non-hierarchical) Kondo model can only be computed numerically.

In this paper, we present a hierarchical version of the Kondo model,
whose renormalization group flow equations can be written out {\it exactly},
with no need for perturbative methods, and show that the flow admits a
non-trivial fixed point.
In this model, the transition from the fixed point can be studied by iterating
an {\it explicit} map,
which allows us to compute reliable numerical values for the {\it Kondo temperature},
that is the temperature at which the Kondo effect emerges, which is related to the
number of iterations required to reach the non-trivial fixed point from the
trivial one. This
temperature has been found to obey the expected scaling relations, as predicted
in \cite{Wi975}.
\*

It is worth noting that the Kondo model (or rather a linearized continuum
version of it) was shown to be exactly solvable by Andrei \cite{An980} 
at
$h=0$, as well as at $h\ne0$, \cite{AFL983},
using Bethe Ansatz, who proved the existence of a Kondo effect in that model.
The aim of the present work is to show how the Kondo effect can be understood
as coming from a non-trivial fixed point in a renormalization group analysis
(in the context of a hierarchical model)
rather than a proof of the existence of the Kondo effect, which has already
been carried out in Ref.\cite{An980,AFL983}.

%%%%%%%%%%%%%%%%%%%%%%%%%%%%%%%%%%%%%%%%%%%%%%%
%%%%%%%%%%%%%%%%%%%%%%%%%%%%%%%%%%%%%%%%%%%%%%%
\section{Kondo model and main results}
\label{sec2}
%%%%%%%%%%%%%%%%%%%%%%%%%%%
%%%%%%%%%%%%%%%%%%%%%%%%%%%

Consider a {\it 1-dimensional} Fermi gas of spin-1/2 ``electrons'', and a
spin-1/2 fermionic ``impurity'' with {\it no} interactions.  It is well
known that: \*
\0(1) the
magnetic susceptibility of the impurity diverges as $\b=\frac1{k_B
  T}\to\infty$ while
\\
(2) both the total susceptibility per particle of the electron gas (\ie the
response to a field acting on the whole sample) \cite{Ki976} and the
susceptibility to a magnetic field acting on a single lattice site of the
chain (\ie the response to a field localized on a site, say at $0$) are
finite at zero temperature (see remark (1) in App.\ref{appG} for a discussion
of the second claim).
\*

The question that will be addressed in this work is whether a small
coupling of the impurity {fermion} with the electron gas can change this
behavior, that is whether the susceptibility of the impurity interacting
with the electrons diverges or not.
To that end we will study a model inspired by the Kondo
Hamiltonian which, expressed in second quantized form, is
\begin{eqnarray} &&H_0=
\sum_{\a\in\{\up,\down\}}\Big(\sum_{x=-{L}/2}^{{L}/2-1}
c^+_\a(x)\,\left(-\frac{\D}2-1\right)\,c^-_\a(x) \Big)
\nn\\
&&H_K=H_0+V_0+V_h\defi H_0+V
\Eq{e2.1}\\
&&V_0=-\l_0\kern-5mm\sum_{j=1,2,3\atop\a_1,\a_2,\a_3,\a_4}
\kern-3mm c^+_{\a_1}(0)\s^j_{\a_1,\a_2}c^-_{\a_2}(0)\,
d^+_{\a_3}\s^j_{\a_3,\a_4}d^-_{\a_4}\nn\\
&&V_h= -h \,\sum_{j=1,2,3}\Bo_j\sum_{(\a,\a')\in\{\up,\down\}^2} d^+_\a \s^j_{\a,\a'} d_{\a'}
\nn
\end{eqnarray}
where $\l_0,h$ are the interaction and magnetic field strengths and
\*
\0(1) $c_\a^\pm(x),d^\pm_\a, \,\a=\uparrow,\downarrow$ are creation and annihilation
operators corresponding respectively to electrons and the
impurity
\\
(2) $\s^j,\,j=1,2,3$, are the Pauli matrices
\\
(3) $x$ is on the unit lattice and
$-{L}/2$, ${L}/2$ are identified (periodic boundary)
\\
(4) $\D f(x)= f(x+1)-2f(x)+f(x-1)$ is the discrete Laplacian.
\\
(5) $\Bo\equiv(\Bo_1,\Bo_2,\Bo_3)$ is a norm-1 vector which specifies the
direction of the magnetic field.
\\
(6) the $-1$ term in $H_0$ is the chemical potential, set to $-1$ (half-filling)
for convenience.\\

The model Eq.\equ{e2.1} differs from the original Kondo model in which
the interaction was
$$-\l_0\sum_{j=1}^3
c^+_{\a_1}(0)\s^j_{\a_1,\a_2}c^-_{\a_2}(0) \,\t^j$$
where $\t^j$ is the
$j$-th Pauli matrix and acts on the spin of the impurity. The two models
are closely related and equivalent for our purposes (see App.\ref{appA}).
The technical advantage of the model
Eq.\equ{e2.1}, is that it allows us set up the problem via a functional
integral to exploit fully the remark that ``since the Kondo problem of
the magnetic impurity treats only a single-point impurity, the question
reduces to a sum over paths in only one (``time'') dimension''
\cite{AYH970b}.
The formulation in Eq.(\ref{e2.1}) was introduced in \cite{An980}.
\*

The model will be said to exhibit a {\it Kondo effect} if, no matter how
small the coupling $\l_0$ is, as long as it is {\it antiferromagnetic} (\ie
$\l_0<0$), the susceptibility {\it remains finite and positive as
  $\b\to\infty$ and continuous as $h\to0$}, while it diverges in presence
of a ferromagnetic (\ie $\l_0>0$) coupling. The soluble model in
\cite{An980} and Wilson's version of the model in Eq.\equ{e2.1} do exhibit
the Kondo effect.
\*

\0{\it Remark}: In the present work, the {\it Kondo effect} is defined as
an effect on the susceptibility of the impurity, and not on the resistivity
of the electrons of the chain, which, we recall, was Kondo's original
motivation \cite{Ko964}. The reason for this is that the magnetic
susceptibility of the impurity is easier to compute than the resistivity of
the chain, but still exhibits a non-trivial effect, as discussed by Wilson
\cite{Wi975}.
\*

Here the same questions will be studied in a hierarchical model defined
below.  The interest of this model is that various observables can be
computed by iterating a map, which is explicitly computed and called
the ``beta function'', involving few (nine) variables, called ``running
couplings''. The possibility of computing the beta function exactly for
general fermionic hierarchical models has been noticed and used in
\cite{Do991}.
\*

\0{\it Remark}: The hierarchical Kondo model {\it will not be an approximation} of
Eq.\equ{e2.1}. It is a model that illustrates a simple mechanism for the
control of the growth of relevant operators in a theory exhibiting a Kondo effect.
\*

The reason why the Kondo effect is not easy to understand is that it is an
intrinsically non-perturbative effect, in that the impurity
susceptibility in the interacting model is qualitatively different from
its non-interacting counterpart. In the sense of the renormalization
group it exhibits several ``relevant'', ``marginal'' and ``irrelevant''
running couplings: this makes any naive perturbative approach hopeless
because all couplings become large (\ie at least of O(1)) at large scale,
no matter how small the interaction is, as long as $\l_0<0$, and thus leave
the perturbative regime.  It is among the simplest cases in which
asymptotic freedom {\it does not occur}. {Using the fact that the beta
  function of the hierarchical model can be computed exactly, its
  non-perturbative regime can easily be investigated.}  \*

In the sections below, we will define the hierarchical Kondo model and show
numerical evidence for the following claims (in principle, such claims
could be proved using computer-assisted methods, though, since the
numerical results are very clear and stable, it may not be worth the
trouble).  \*

{\it If the interactions between the electron spins and the impurity are
  antiferromagnetic } (\ie $\l_0<0$ in our notations), then

\0(1) The {\it existence of a Kondo effect} can be proved in spite of the
lack of asymptotic freedom and formal growth of the effective Hamiltonian
away from the trivial fixed point, {\it because the beta function can be
  computed exactly} (in particular non-pertubatively).

\0(2) In addition, there exists an inverse temperature
  $\b_K=2^{n_K(\l_0)}$ called the {\it Kondo} inverse temperature, such
  that the Kondo effect manifests itself for $\b>\b_K$. Asymptotically as
  $\l_0\to0$, $n_K(\l_0)=c_1|\l_0|^{-1}+O(1)$.

\0(3) It will appear that perturbation theory can only work to describe
properties measurable up to a length scale $2^{n_2(\l_0)}$, in which
$n_2(\l_0)$ depends on the coupling $\l_0$ between the impurity and the
electron chain and, asymptotically as $\l_0\to0$,
$n_2(\l_0)=c_2\log|\l_0|^{-1}+O(1)$ for some $c_2>0$; at larger scales
perturbation theory breaks down and the evolution of the running
couplings is controlled by a non-trivial fixed point (which can be computed
exactly).

\0(4) Denoting the magnetic field by $h$, if $h>0$ and $\b_Kh\ll1$, the
flow of the running couplings tends to a trivial fixed point
($h$-independent but different from $0$) which is reached on a scale $r(h)$
which, {asymptotically} as $h\to0$, is $r(h)=c_r \log h^{-1} +O(1)$.  \*

{\it The picture is completely different in the ferromagnetic case,} in
which the susceptibility diverges at zero temperature and the flow of the
running couplings is not controlled by the non trivial fixed point.  \*

\0{\it Remark}: Unlike in the model studied by Wilson \cite{Wi975}, the $T=0$
nontrivial fixed point is {\it not} infinite in the hierarchical Kondo
model: this shows that the Kondo
effect can, in some models, be somewhat subtler than a rigid locking of
the impurity spin with an electron spin\cite{No974}.
\*

Technically this is one of the few cases in which functional
integration for fermionic fields is controlled by a non-trivial fixed
point and can be performed rigorously and applied to
a concrete problem.
\*
\0{\it Remark:} (1) It is worth stressing that in a system consisting of
two classical spins with coupling $\l_0$ the susceptibility at $0$ field is
$4\b(1+e^{-2\b\l_0})^ {-1}$, hence it vanishes at $T=0$ in the
antiferromagnetic case and diverges in the ferromagnetic and in the free
case. Therefore this simple model does not exhibit a Kondo effect.
\\
(2) In the exactly solvable XY model, which can be shown to be equivalent
to a spin-less analogue of Eq.\equ{e2.1}, the susceptibility can
be shown to diverge in the $\b\to\infty$ limit, see
App.\ref{appG}, \ref{appH} (at least for some boundary conditions).
Therefore this model does not exhibit a Kondo effect either.

%%%%%%%%%%%%%%%%%%%%%%%%%%%%%%%%%%%%%%%%%%%%%%%
%%%%%%%%%%%%%%%%%%%%%%%%%%%%%%%%%%%%%%%%%%%%%%%
\section{Functional integration in the Kondo model}
\label{sec3}
%%%%%%%%%%%%%%%%%%%%%%%%%%%
%%%%%%%%%%%%%%%%%%%%%%%%%%%

In \cite{Wi975}, Wilson studies the Kondo problem using renormalization
group techniques in a Hamiltonian context. In the present work, our aim is
to reproduce, in a simpler model, analogous results using a formalism based
on functional integrals.

In this section, we give a rapid review of the functional integral
formalism we will use, following \cite{BG990,Sh994}. We will not attempt to
reproduce all technical details, since it will merely be used
as an inspiration for the definition of the hierarchical model in
section~\ref{sec4}.

We introduce an extra dimension, called {\it imaginary time}, and define
new creation and annihilation operators:
\be
c_{\a}^\pm(x,t)\defi e^{tH_0}c_{\a}^\pm(x) e^{-tH_0},\quad
d^\pm_{\a}(t)\defi e^{tH_0}d^\pm_{\a}e^{-tH_0},
\Eq{e3.1}\ee
for $\a\in\{\up,\down\}$, to which we associate anti-commuting
{\it Grassmann variables}:
\be
c_\alpha^\pm(x,t)\longmapsto\ps_\alpha^\pm(x,t),\quad
d_\alpha^\pm(t)\longmapsto\f_\alpha^\pm(t).
\Eq{e3.2}\ee
Functional integrals are expressed as ``Gaussian integrals'' over the
Grassmann variables:\footnote{\small This means that all integrals will be
  defined and evaluated via the ``Wick rule''.}
\be
\int P(d\f)P(d\ps)\cdot\ \defi\ \int \prod_\a P(d\f_\a) P(d\ps_\a)\ \cdot \Eq{e3.3}\ee
$P(d\f_\a)$ and $P(d\ps_\a)$ are Gaussian measures whose covariance
(also called {\it propagator}) is defined by
\begin{eqnarray}
g_{\ps,\a}(x-x',t-t')
&\defi&
\left\{\begin{array}{ll}
\displaystyle\frac{\hbox{Tr}\,
e^{-\b H_0}c^-_{\a}(x,t)c^+_{\a}(x',t')}{\hbox{Tr}\,e^{-\b H_0}}
&\mathrm{if\ } t>t'\\[0.5cm]
\displaystyle-\frac{\hbox{Tr}\,e^{-\b H_0}
c^+_{\a}(x',t')c^-_{\a}(x,t)}{\hbox{Tr}\,e^{-\b H_0}}
&\mathrm{if\ } t\le t'
\end{array}\right.
\Eq{e3.4}\\[0.5cm]
g_{\f,\a}(t-t')&\defi&
\left\{\begin{array}{ll}
\hbox{Tr}\,d^-_{\a}(t)d^+_{\a}(t')
&\mathrm{if\ } t>t'\\[0.5cm]
-\hbox{Tr}\,d^+_{\a}(t')d^-_{\a}(t)
&\mathrm{if\ }t\le t'
\end{array}\right..\nn
\end{eqnarray}
By a direct computation \cite{BG990}, Eq.(2.7), we find that in the limit
$L,\b\to\infty$, if $e(k)\defi(1-\cos k) -1\equiv -\cos k$ (assuming the
Fermi level is set to $1$, \ie the Fermi momentum to $\pm\frac\p2$) then
\be
g_{\ps,\a}(\x,\t)  =\int\frac{dk_0 dk}{(2\p)^2}\,{e^{-ik_0(\t+0^-)-ik\x}
\over-ik_0+e(k) },\quad
g_{\f,\a}(\x,\t) = \int\frac{dk_0}{2\p}\,{e^{-ik_0(\t+0^-)}
\over-ik_0}.
\Eq{e3.5}\ee
If $\b,L$ are finite, $\int\,\frac{dk_0 dk}{(2\p)^2}$ in Eq.\equ{e3.5}
  has to be understood as $\frac1\b \sum_{k_0} \frac1L \sum_k$, where $k_0$
  is the ``Matsubara momentum'' $k_0= \frac\p{\b} +\frac{2\p}\b n_0$,
  $n_0\in\mathbb Z$, $|n_0|\le\frac12\b$, and $k$ is the linear momentum
  $k=\frac{2\p}L n$, $n\in [-L/2,L/2-1]\cap\mathbb Z$.

In the functional representation, the operator $V$ of Eq.\equ{e2.1} is
substituted with the following function of the Grassmann variables
\equ{e3.2}:
\begin{eqnarray}
V(\ps,\f)&=&
-h \, \sum_{j\in\{1,2,3\}}\Bo_j\int dt\sum_{(\a,\a')\in\{\up,\down\}^2} \f^+_{\a}\s^j_{\a,\a'} \f^-_{\a'}\Eq{e3.6}\\
&&-\l_0\sum_{{j\in\{1,2,3\}}\atop{\a_1,\a'_1,\a_2,\a_2'\in
\{\uparrow,\downarrow\}}}\int dt
(\ps^+_{\a_1}(0,t)\sigma^j_{\a_1,\a'_1} \ps^-_{\a'_1}(0,t))
(\f^+_{\a_2}(t)\sigma^j_{\a_2,\a_2'} \f^-_{\a_2'}(t)).\nn
\end{eqnarray}
Notice that $V$ only depends on the fields located at the site $x=0$.
This is important because it will allow us to reduce the problem to
a 1-dimensional one \cite{AY969,AYH970b}.

The average of a physical observable $F$ localized at $x=0$, which is a
polynomial in the fields $\ps_{\a}^\pm(0,t)$ and $\f_\a^\pm(t)$, will be
denoted by
\be
\media{F}_K\defi\frac1Z\int P(d\f)P_0(d\ps)\,
e^{-V(\ps,\f)} \,F,\quad
\Eq{e3.7}\ee
in which $P_0(d\ps)$ is the Gaussian Grassmannian measure over the fields
$\ps_\a^\pm(0,t)$ localized at the site $0$ and with propagator
$g_{\ps,\a}(0,\tau)$ and $Z$ is a normalization factor.

The propagators can be split into scales by introducing a smooth cutoff
function $\chi$ which is different from $0$ only on $(\frac14,1)$ and,
denoting $N_\b\defi \log_2 \b$, is such that
$\sum_{m=-N_\b}^\infty\chi(2^{-2m}z^2)=1$ for all
$|z|\in[\frac{\p}{\b},N_\b]$. Let
\begin{eqnarray}
g_{\ps}^{[m]}(0,\tau)&\defi&\sum_{\o\in\{-,+\}}\int\frac{dk_0
dk}{(2\pi)^2}
{e^{-ik_0(\tau+0^-)}
\over-ik_0+e(k)}\chi(2^{-2m}((k-\o\pi/2)^2+k_0^2))\nn\\
g_\ps^{[\mathrm{uv}]}(0,\tau)&\defi&
g_\ps(0,\tau)-\sum_{m=-N_\beta}^{m_0}
g_{\ps}^{[m]}(0,\tau)
\Eq{e3.8}\\
g_\f^{[m]}(\tau)&\defi&
\int\frac{dk_0}{2\pi}\,{e^{-ik_0(\tau+0^-)}
\over-ik_0}\chi(2^{-2m}k_0^2)\nn\\
g_\f^{[\mathrm{uv}]}(\tau)&\defi&
g_\f(\tau)-\sum_{m=-N_\beta}^{m_0}g_\f^{[m]}(\tau).
\nn
\end{eqnarray}
where $m_0$ is an integer of order one (see below).
\*
\noindent{\it Remark}: The $\o=\pm$ label refers to the ``quasi
particle'' momentum $\o p_F$, where $p_F$ is the Fermi momentum. The
usual approach \cite{BG990,Sh994} is to decompose the field $\ps$ into
quasi-particle fields:
\be
\ps^\pm_{\a}(0,t)=\sum_{\o=\pm} \ps^\pm_{\o,\a}(0,t),
\Eq{e3.9}
\ee
indeed, the introduction of quasi particles \cite{BG990,Sh994}, is key to
separating the oscillations on the Fermi scale $p_F^{-1}$ from the
propagators thus allowing a ``naive'' renormalization group analysis of
fermionic models in which multiscale phenomena are important (as in the
theory of the ground state of interacting fermions \cite{BG990,BGPS994},
or as in the Kondo model). In this case, however, since the fields are
evaluated at $x=0$, such oscillations play no role, so we will not decompose
the field.
\*

We set $m_0$ to be small enough (\ie negative enough) so that
$2^{m_0}p_F\le1$ and introduce a first {\it approximation}: we neglect
$g_{\ps}^{[\mathrm{uv}]}$ and $g_\f^{[\mathrm{uv}]}$, and replace $e(k)$ in
Eq.\equ{e3.5} by its first order Taylor expansion around $\o p_F$, that is
by $\o k$. As long as $m_0$ is small enough, for all $m\le m_0$ the
supports {of the two functions $\chi(2^{-2m}((k-\o\pi/2)^2+k_0^2))$, $\o=\pm1$}, which appear
in the first
of Eqs.\equ{e3.8} do not intersect, and approximating $e(k)$ by $\o k$ is
reasonable.  We shall hereafter fix $m_0=0$ thus avoiding the introduction
of a further length scale and keeping only two scales when no impurity is
present.

Since we are interested in the {\it infrared} properties of the system, we
consider such approximations as minor and more of a {\it simplification}
rather than an approximation, since the ultraviolet regime is expected to
be trivial because of the discreteness of the model in the operator
representation.

After this approximation, the propagators of the model reduce to
\begin{samepage}\postdisplaypenalty0
\begin{eqnarray}
g_{\ps}^{[m]}(0,\tau)&=&
\sum_{\o\in\{-,+\}}\int\frac{dk_0 dk}{(2\pi)^2}
{e^{-ik_0(\tau+0^-)}
\over-ik_0+\o k}\chi(2^{-2m}(k^2+k_0^2))\nn\\
g_\f^{[m]}(\tau)&=&
\int\frac{dk_0}{2\pi}\,{e^{-ik_0(\tau+0^-)}
\over-ik_0}\chi(2^{-2m}k_0^2).
\Eq{e3.10}\end{eqnarray}
\end{samepage}
and satisfy the following {\it scaling} property:
\be\kern-3mm
g_{\ps}^{[m]}(0,\tau)=2^{m}g_{\ps}^{[0]}(0,2^{m}\tau),\quad
g_\f^{[m]}(\tau)=g_\f^{[0]}(2^{m}\tau).
\Eq{e3.11}\ee
The Grassmannian fields are similarly decomposed into scales:
\be
\ps^\pm_\a(0,t)=
\sum_{m=-N_\beta}^{0} 2^{\frac{m}2}\ps_{\a}^{[m]\pm}(0,2^{-m}t),\quad
\f^\pm_\a(t)=\sum_{m=-N_\beta}^{0} \kern-2mm \f_a^{[m]\pm}(2^{-m}t)
\Eq{e3.12}\ee
with $\ps_{\a}^{[m]}(0,t)$ and $\f_\a^{[m]}(t)$ being, respectively,
assigned the following propagators:
\begin{eqnarray}
\int P_0(d\ps^{[m]})\ps_{\a}^{[m]-}(0,t)\ps_{\a'}^{[m]+}(0,t')
&\defi&\delta_{\a,\a'}g_{\ps}^{[0]}(0,2^{m}(t-t'))\nn\\
\int P(d\f^{[m]})\f_{\a}^{[m]-}(t)\f_{\a'}^{[m]+}(t')
&\defi&\delta_{\a,\a'}g_{\f}^{[0]}(2^{m}(t-t')).
\Eq{e3.13}\end{eqnarray}
{\it Remark:} by Eq.\equ{e3.11} this is equivalent to stating that the
propagators associated with the $\ps^{[m]},\f^{[m]}$ fields are
$2^{-m}g^{[m]}$ and $g^{[m]}$, respectively.
\*
Finally, we define
\be
\ps_{\a}^{[\le m]\pm}(0,t)\defi\sum_{m'=-N_\beta}^{m}
2^{\frac{m'}2}\ps_{\a}^{[m']\pm}(0,t),\quad
\f_{\a}^{[\le
  m]\pm}(t)\defi\sum_{m'=-N_\beta}^{m}\f_{\a}^{[m']\pm}(t).
\Eq{e3.14}\ee
Notice that the functions $g_\ps^{[m]}(\xi,\tau),g_\f^{[m]}(\tau)$ decay
faster than any power as $\tau$ tends to $\infty$ (as a consequence of the
smoothness of the cut-off function $\chi$), so that at any fixed scale
$m\le 0$, fields $\ps^{[m]},\f^{[m]}$ that are separated in time by more
than $2^{-m}$ can be regarded as (almost) independent.

The decomposition into scales allows us to express the quantities
in Eq.\equ{e3.7} inductively (see (\ref{e3.16})).
For instance the partition function $Z$ is given by
\be
Z=\exp\Big(-\beta\sum_{m=-N_\beta}^{0} c^{[m]}\,\Big)
\Eq{e3.15}\ee
where, for $N_\beta<m\le 0$,
\begin{samepage}\postdisplaypenalty0
\begin{eqnarray}
\beta c^{[m-1]}+V^{[m-1]}(\ps^{[\le m-1]},\f^{[\le m-1]})
&\defi&
-\log\int P(d\ps^{[m]})P(d\f^{[m]})\,e^{-V^{[m]}(\ps^{[m]},\f^{[m]})}\nn\\
 V^{[0]}(\ps^{[\le0]},\f^{[\le0]})&\defi&
V(\ps^{[\le0]},\f^{[\le0]})
\Eq{e3.16}\end{eqnarray}
\end{samepage}
in which $c^{[m-1]}\in R$ and $V^{[m-1]}$ has no constant term, \ie no
fields independent term.
%

%%%%%%%%%%%%%%%%%%%%%%%%%%%%%%%%%%%%%%%%%%%%%%%
%%%%%%%%%%%%%%%%%%%%%%%%%%%%%%%%%%%%%%%%%%%%%%%
\section{Hierarchical Kondo model}
\label{sec4}
%%%%%%%%%%%%%%%%%%%%%%%%%%%
%%%%%%%%%%%%%%%%%%%%%%%%%%%

In this section, we define a hierarchical Kondo model, localized at $x=0$ (the
location of the impurity), inspired by the discussion in the previous section and
the remark that the problem of the Kondo effect is reduced there to the
evaluation of a functional integral over the fields $\ps(x,t),\f(t)$ {\it
with $x\equiv0$}. The hierarchical model is a model that is represented using
a functional integral, that shares a few features with the functional integral
described in Sec.\ref{sec3}, which are essential to the Kondo effect. Therein the fields $\ps^{[m]}$ and $\f^{[m]}$ evaluated at
$x=0$ are assumed to be constant in $t$ on scale $2^{-m},
m=0,-1,-2\ldots $, and the propagators $g_{\ps}^{[m]}(0,\tau)$ and
$g_\f^{[m]}(\tau)$ with large Matsubara momentum $k_0$
are neglected ($g^{[uv]}=0$ in Eq.\equ{e3.8}).

The hierarchical Kondo model is defined by introducing a family of
{\it hierarchical fields} and specifying a {\it propagator} for
each pair of fields. The average of any monomial of fields is then
computed using the Wick rule.

As a preliminary step, we pave the time axis $R$ with boxes of size
$2^{-m}$ for every $m\in\{0,-1,\ldots,-N_\beta\}$. To that end, we
define the set of {\it boxes on scale $m$} as
\be
{\mathcal Q}_m\defi\left\{[i  2^{|m|}, (i+1) 2^{|m|})\right\}_{i=0,1,\cdots,2^{N_\beta-|m|}-1,\atop m=0,-1,\ldots\kern1.4cm}
\Eq{e4.1}\ee
Given a box $\Delta\in{\mathcal Q}_m$, we define $t_\Delta$ as the center
of $\Delta$; conversely, given a point $t\in R$, we define
$\Delta^{[m]}(t)$ as the (unique) box on scale $m$ that contains $t$.

A naive approach would then be to define the hierarchical model in terms of
the fields $\ps^{[m]}_{t_\Delta}$ and $\f^{[m]}_{t_\Delta}$, and neglect
the propagators between fields in different boxes, but, as we will see
below, such a model would be trivial (all propagators would vanish because
of Fermi statistics).

Instead, we further decompose each box into two {\it half boxes}: given
$\Delta\in{\mathcal Q}_m$ and $\eta\in\{-,+\}$, we define
\be
\Delta_{\eta}\defi\Delta^{[m+1]}(t_{\Delta}+\eta2^{-m-2})
\Eq{e4.2}\ee
for $m< 0$ and similarly for $m=0$. Thus $\Delta_{-}$ is the lower half
of $\Delta$ and $\Delta_{+}$ the upper half.

The elementary fields used to define the hierarchical Kondo model will be
{\it constant on each half-box} and will be denoted by
$\ps_\a^{[m]\pm}(\Delta_{\eta})$ and $\f_{\a}^{[m]\pm}(\Delta_{\eta})$
for $m\in\{0,-1,\cdots,$ $-N_\beta\}$, $\Delta\in\mathcal Q_m$,
$\eta\in\{-,+\}$, $\a\in\{\uparrow,\downarrow\}$.  \*

We now define the propagators associated with $\ps$ and $\f$. The idea is
to define propagators that are {\it similar} \cite{Wi965,Wi970,Dy969}, in
a sense made more precise below, to the non-hierarchical propagators
defined in Eq.\equ{e3.4}. Bearing that in mind, we compute the value of the
non-hierarchical propagators between fields at the centers of two half
boxes: given a box $\Delta\in\mathcal Q_{0}$ and $\eta\in\{-,+\}$, let
$\delta\defi2^{-1}$ denote the distance between the centers of $\Delta_-$ and
$\Delta_+$, we get
\begin{samepage}\postdisplaypenalty0
\begin{eqnarray}
g^{[0]}_{\ps}(0,\eta\delta)&=&\eta\sum_{\o=\pm}\int \frac{dkdk_0}{(2\pi)^2}
\frac{k_0\sin(k_0\delta)}{k_0^2+k^2}\chi(k^2+k_0^2)\defi\eta a\nn\\
g^{[0]}_{\f}(\eta\delta)&=&\eta\int\frac{dk_0}{2\pi}
\frac{\sin(k_0\delta)}{k_0}\chi(k_0^2)\,\defi\,\eta b
\Eq{e4.3}
\end{eqnarray}
\end{samepage}
in which $a$ and $b$ are constants, see \cite[p.4465]{AYH970b}. We
define the hierarchical propagators, drawing inspiration from
Eq.\equ{e4.3}. In an effort to make computations more explicit, we set
$a=b\equiv1$ and define
\be
\left<\ps_{\a}^{[m]-}(\Delta_{-\eta})\ps_{\a}^{[m]+}(\Delta_{\eta})\right
>\defi \eta,\quad
\left<\f_{\a}^{[m]-}(\Delta_{-\eta})\f_{\a}^{[m]+}(\Delta_{\eta})\right>
\defi \eta
\Eq{e4.4}\ee
for $m\in\{0,-1,\cdots,-N_\beta\}$, $\eta\in\{-,+\}$, $\Delta\in\mathcal
Q_m$, $\a\in\{\downarrow,\uparrow\}$. All other propagators
are $0$. Note that if we had not defined the model using half boxes, all
the propagators in Eq.\equ{e4.3} would vanish, and the model would be
trivial.  \*

In order to link back to the non-hierarchical model, we define the
  following quantities: for all $t\in R$,
\be
\ps^{\pm}_{\a}(0,t)\defi
\sum_{m=-N_\beta}^{0}2^{\frac{m}2}
\ps^{[m]\pm}_{\a}(\Delta^{[m+1]}(t)),\quad
\f^{\pm}_{\a}(t)\defi\sum_{m=-N_\beta}^{0}
\f^{[m]\pm}_{\a}(\Delta^{[m+1]}(t))
\Eq{e4.5}\ee
(recall that $m\le0$ and $\Delta^{[m]}(t)\supset\Delta^{[m+1]}(t)$). The
hierarchical model for the on-site Kondo effect so defined is such that the
propagator on scale $m$ between two fields vanishes unless both fields belong
to the
same box and, at the same time, to two different halves within that box. In
addition, given $t$ and $t'$ that are such that $|t-t'|>2^{-1}$, there
exists one and only one scale $m_{(t-t')}$ that is such that
$\Delta^{[m_{(t-t')}]}(t)=\Delta^{[m_{(t-t')}]}(t')$ and
$\Delta^{[m_{(t-t')}+1]}(t)\neq\Delta^{[m_{(t-t')}+1]}(t')$. Therefore
$\forall(t,t')\in R^2$, $\forall(\a,\a')\in\{\uparrow,\downarrow\}^2$,
\be
\media{\ps_\a^-(0,t)\ps_{\a'}^+(0,t')}=\delta_{\a,\a'}
2^{m_{(t-t')}}\mathrm{sign}(t-t').
\Eq{e4.6}\ee
The non-hierarchical analog of Eq.\equ{e4.6} is (we recall that
$\media{\cdot}_K$ was defined in Eq.\equ{e3.7})
\be
\media{\ps_\a^-(0,t)\ps_{\a'}^+(0,t')}_K
=\delta_{\a,\a'}
\sum_{m'=-N_\beta}^{0}2^{m'}g_{\ps}^{[0]}(0,2^{m'}(t-t'))
\Eq{e4.7}\ee
from which we see that the hierarchical model boils down to neglecting the
$m'$ that are ``wrong'', that is those that are different from
$m_{(t-t')}$, and approximating $g^{[m]}_{\ps}$ by
$\mathrm{sign}(t-t')$. Similar considerations hold for $\f$.  \*

The physical observables $F$ considered here will be
polynomials in the hierarchical fields;
their averages, by analogy with Eq.\equ{e3.7}, will be
\be
\frac1Z\media{e^{-V(\ps,\f)}F},\quad
Z=\media{e^{-V(\ps,\f)}}
\Eq{e4.8}\ee
(in which $\media{\cdot}$ is computed using the Wick rule and Eq.\equ{e4.4})
and, similarly to Eq.\equ{e3.6},
\begin{eqnarray}
V(\ps,\f)&=&
-h \, \sum_{j\in\{1,2,3\}}\Bo_j\int dt\sum_{(\a,\a')\in\{\up,\down\}^2} \f^+_{\a}\s^j_{\a,\a'} \f^-_{\a'}\Eq{e4.9}\\
&&-\l_0\sum_{{j\in\{1,2,3\}}\atop{\a_1,\a'_1,\a_2,\a_2'\in
\{\uparrow,\downarrow\}}}\int dt
(\ps^+_{\a_1}(0,t)\sigma^j_{\a_1,\a'_1} \ps^-_{\a'_1}(0,t))
(\f^+_{\a_2}(t)\sigma^j_{\a_2,\a_2'} \f^-_{\a_2'}(t)).\nn
\end{eqnarray}
in which $\ps^\pm_\a(0,t)$ and $\f^\pm_\a(t)$ are now defined in Eq.\equ{e4.5}.
\*

{Note that since the model defined above only involves field localized at
the impurity site, that is at $x=0$, we only have to deal with $1$-dimensional
fermionic fields.}
{\it This does not mean} that the lattice
supporting the electrons plays no role: on the contrary it will show up,
and in an essential way, because the ``dimension'' of the electron field
will be different from that of the impurity, as made already manifest by
the factor $2^m\tende{m\to-\infty}0$ in
Eq.\equ{e4.6}.
\*

Clearly several properties of the non-hierarchical propagators,
Eq.\equ{e3.10}, are not reflected in Eq.\equ{e4.6}. However it will be seen
that even so simplified the model exhibits a ``Kondo effect'' in the sense
outlined in Sec.\ref{sec1}.

%%%%%%%%%%%%%%%%%%%%%%%%%%%%%%%%%%%%%%%%%%%%%%%
%%%%%%%%%%%%%%%%%%%%%%%%%%%%%%%%%%%%%%%%%%%%%%%
\section{Beta function for the partition function.}
\label{sec5}
%%%%%%%%%%%%%%%%%%%%%%%%%%%
%%%%%%%%%%%%%%%%%%%%%%%%%%%

In this section, we show how to compute the partition function $Z$ of the
hierarchical Kondo model (see Eq.\equ{e4.8}), and introduce the concept of
a {\it renormalization group flow} in this context.
We will first restrict the discussion to the $h=0$ case, in which
$V=V_0$; the case $h\ne0$ is discussed in Sec.\ref{sec6}.

The computation is carried out in an inductive fashion by splitting the
average $\media{e^{V_0(\ps,\f)}}$ into partial averages over the fields on
scale $m$. Given $m\in\{0,-1,\cdots,-N_\b\}$, we define $\media{\cdot}_m$
as the partial average over $\ps^{[m]\pm}_{\a}(\D_\eta)$ and
$\f^{[m]\pm}_\a(\D_\eta)$ for $\a\in\{\uparrow,\downarrow\}$,
$\D\in\mathcal Q_m$ and $\eta\in\{-,+\}$, as well as
\be
\ps_{\a}^{[\le m]\pm}(\D_\eta)\defi\frac1{\sqrt2}
\ps_{\a}^{[\le m-1]\pm}(\D)+\ps_{\a}^{[m]\pm}(\D_\eta),\quad
\f_{\a}^{[\le m]\pm}(\D_\eta)\defi\f_{\a}^{[\le m-1]\pm}(\D)
+\f_{\a}^{[m]\pm}(\D_\eta)
\Eq{e5.1}\ee
and for $\D\in\mathcal Q_{-m},\,m<-N_\b$,
\be
\ps_{\a}^{[\le m]}(\D_\eta)\defi0,\qquad
\f_{\a}^{[\le m]}(\D_\eta)\defi0.
\Eq{e5.2}\ee
Notice that the fields $\ps_{\a}^{[\le m-1]\pm}(\D)$ and $\f_{\a}^{[\le
    m-1]\pm}(\D)$ play (temporarily) the role of {\it external fields} as
they do not depend on the index $\eta$, and are therefore independent of
the half box in which the {\it internal fields} $\ps_{\a}^{[\le
    m]\pm}(\D_\eta)$ and $\f_{\a}^{[\le m]\pm}(\D_\eta)$ are defined.  In
addition, by iterating Eq.\equ{e5.1}, we can rewrite Eq.\equ{e4.5} as
\be\kern-3mm \ps_\a^\pm(t)\equiv\ps_{\a}^{[\le0]\pm}(\D^{[1]}(t)),\quad
\f_\a^\pm(t)\equiv\f_\a^{[\le0]\pm}(\D^{[1]}(t))
\Eq{e5.3}\ee
We then define, for $m\in\{0,-1,\cdots,-N_\b\}$,
\begin{samepage}\postdisplaypenalty0
\begin{eqnarray}
\b c^{[m]}+V^{[m-1]}(\ps^{[\le m-1]},\f^{[\le m-1]})
&\defi&
-\log\media{e^{-V^{[m]}(\ps^{[\le m]},\f^{[\le m]})}}_m\nn\\
V^{[0]}(\ps^{[\le0]},\f^{[\le 0]})&\defi&V_0(\ps^{[\le0]},\f^{[\le 0]})
\Eq{e5.4}\end{eqnarray}
\end{samepage}
in which $c^{[m-1]}\in R$ is a constant and $V^{[m-1]}$ contains no
constant term. By a straightforward induction, we then find that
$Z$ is given again by Eq.\equ{e3.15} with the present definition of $c^{[m]}$
 (see Eq.\equ{e5.4}).
\*

We will now prove by induction that the hierarchical Kondo model defined
above is {\it exactly solvable}, in the sense that Eq.\equ{e5.4} can be written
out {\it explicitly} as a {\it finite} system of equations.  To that end it
will be shown that $V^{[m]}$ can be parameterized by only four real
numbers, $\Ba^{[m]}=(\alpha^{[m]}_0,\cdots,\alpha^{[m]}_3)\in R^4$ and, in
the process, the equation relating $\Ba^{[m]}$ and
$\Ba^{[m-1]}$ (called the {\it beta function}) will be computed:
\be
-V^{[m]}(\ps^{[\le m]},\f^{[\le m]})
=\sum_{\Delta\in\mathcal
  Q_m}\sum_{n=0}^3\alpha_n^{[m]}
\sum_{\h=\pm}O_{n,\eta}^{[\le m]}(\D)
\Eq{e5.5}\ee
where
\begin{eqnarray}
O^{[\le m]}_{0,\eta}(\D)&\defi&\frac12\V A^{[\le m]}_\eta(\D)
\cdot\V B^{[\le m]}_\eta(\D)\nn\\
O^{[\le m]}_{1,\eta}(\D)&\defi&\frac12\V A^{[\le m]}_\eta(\D)^2\nn\\
 O^{[\le m]}_{2,\eta}(\D)&\defi&
 \frac12\V B^{[\le m]}_\eta(\D)^2
\Eq{e5.6}\\
O^{[\le m]}_{3,\eta}(\D)&\defi&\frac12\V A^{[\le m]}_\eta(\D)^2
\V B^{[\le m]}_\eta(\D)^2
\nn\end{eqnarray}
in which $\V A^{[\le m]}$ and $\V B^{[\le m]}$ are vectors of polynomials
in the fields, whose $j$-th component for $j\in\{1,2,3\}$ is
\begin{samepage}\postdisplaypenalty0
\begin{eqnarray}
A^{[\le m]j}_\eta(\D)
&\defi&\sum_{(\a,\a')\in\{\uparrow,\downarrow\}^2}
\ps_{\a}^{[\le m]+}(\D_\eta)\s^j_{\a,\a'}
\ps_{\a'}^{[\le m]-}(\D_\eta)\nn\\
B^{[\le m]j}_\eta(\D)
&\defi&\sum_{(\a,\a')\in\{\uparrow,\downarrow\}^2}
\f_{\a}^{[\le m]+}(\D_\eta)\s^j_{\a,\a'}\f_{\a'}^{[\le m]-}(\D_\eta).
\Eq{e5.7}
\end{eqnarray}
\end{samepage}
For $m=0$, by injecting Eq.\equ{e5.3} into Eq.\equ{e4.9}, we find that
$V^{[0]}$ can be written as in Eq.\equ{e5.5} with $\Ba^{[0]}=(\l_0,0,0,0)$.
As follows from Eq.(\ref{e5.13}) below, for all initial conditions,
the running couplings $\alpha^{[m]}$ remain bounded, and are attracted by
a sphere whose radius is independent of the initial data.

We then compute $V^{[m-1]}$ using Eq.\equ{e5.4} and show it can be written
as in Eq.\equ{e5.5}. We first notice that the propagator in Eq.\equ{e4.4}
is diagonal in $\D$, and does not depend on the value of
$\D$, therefore, we can split the averaging over $\ps^{[m]}(\D_\pm)$
for different $\D$, as well as that over $\f^{[m]}(\D)$. We
thereby find that
\be
\media{e^{\sum_{\D} \sum_{n,\eta}\alpha_n^{[m]}O_{n,\D}^{[\le m]}}}_m
=\prod_{\D}\media{e^{\sum_{n,\eta} \alpha_n^{[m]}
O_{n,\eta}^{[\le m]}(\D)}}_m
\Eq{e5.8}\ee
In addition, we rewrite
\begin{samepage}\postdisplaypenalty0
\begin{eqnarray}
e^{\sum_{n,\eta}\alpha_n^{[m]}O_{n,\eta}^{[\le m]}(\D)}
&=&\prod_{\eta=\pm}e^{\sum_n \alpha_n^{[m]}O_{n,\eta}^{[\le m]}(\D)}\nn\\
&=&\prod_{\eta=\pm}\sum_{k=0}^2\frac1{k!}\Big(\sum_{n=0}^3
\alpha_n^{[m]}O_{n,\eta}^{[\le m]}(\D)\Big)^k
\Eq{e5.9}\end{eqnarray}
\end{samepage}
in which the sum over $k$ only goes up to 2 as a consequence of the
anticommutation relations, see
lemma~\ref{Olemma}; this also allows us to rewrite
\be
\sum_{k=0}^2\frac1{k!}\Big(\sum_{n=0}^3\alpha_n^{[m]}
O_{n,\eta}^{[\le m]}(\D)\Big)^k
=1+\sum_{n=0}^3\ell_n^{[m]}
O_{n,\eta}^{[\le m]}(\D)
\Eq{e5.10}\ee
where
\be
\ell_0^{[m]}=\alpha_0^{[m]},\quad
\ell_1^{[m]}=\alpha_1^{[m]},\quad
\ell_2^{[m]}=\alpha_2^{[m]},\quad
\ell_3^{[m]}=\alpha_3^{[m]}-\frac1{12}(\ell_0^{[m]})^2-\frac12\ell_1^{[m]}\ell_2^{[m]}.
\Eq{e5.11}\ee
At this point, we insert Eq.\equ{e5.10} into Eq.\equ{e5.9} and compute the
average, which is a somewhat long computation, although finite (see
App.\ref{appB} for the main shortcuts).  We find that
\be
\Big\langle
\prod_{\eta=\pm}\Big(1+\sum_{n=0}^3\ell_n^{[m]}O_{n,\eta}^{[\le m]}(\D)\Big)
\Big\rangle_m
=C^{[m]}\Big(1+\sum_{n=0}^3\ell^{[m-1]}_{n}O_{n}^{[\le m-1]}(\D)\Big)
\Eq{e5.12}\ee
with (in order to reduce the size of the following equation, we dropped all
$^{[m]}$ from the right side)
\begin{eqnarray}
C^{[m]}&=&1+
3\ell_0^2+9\ell_1^2+9\ell_2^2+324\ell_3^2\nn\\
\ell_0^{[m-1]}&=&\frac1{C^{[m]}}\Big(\ell_0
+18\ell_0\ell_3+3 \ell_0\ell_2+3 \ell_0\ell_1
-2\ell_0^2\Big)
\nn\\
\ell_1^{[m-1]}&=&\frac1{C^{[m]}}\Big(
\frac12\ell_1+9\ell_2\ell_3
+\frac14\ell_0^2\Big)
\Eq{e5.13}\\
\ell_2^{[m-1]}&=&
\frac1{C^{[m]}}\Big(2\ell_2+36\ell_1\ell_3+ \ell_0^2\Big)\nn\\
\ell_3^{[m-1]}&=&
\frac1{C^{[m]}}\Big(\frac12\ell_3+\frac14\ell_1\ell_2+\frac1{24} \ell_0^2\Big).
\nn\end{eqnarray}
The $\Ba^{[m-1]}$ could then be reconstructed from Eq.\equ{e5.13} by inverting
the map $\Ba\mapsto\Bell$ (see Eq.\equ{e5.11}).
It is nevertheless convenient
to work with the $\Bell$'s as running couplings rather than with the $\Ba$'s.

This concludes the proof of Eq.\equ{e5.5}, and provides an explicit map,
defined in Eq.\equ{e5.13} and which we denote by $\mathcal R$, that is such
that $\Bell^{[m]}=\mathcal R^{|m|}\Bell^{[0]}$. Finally,
the $c^{[m]}$ appearing in Eq.\equ{e5.4} is given by
\be
c^{[m]}=-2^{N_\b+m}\log(C^{[m]})
\Eq{e5.14}\ee
which is well defined: it follows from Eq.\equ{e5.13} that $C^{[m]}\ge1$.
\*

The dynamical system defined by the map $\mathcal R$ in Eq.\equ{e5.13}
admits a few non trivial fixed points. A numerical analysis shows that if
the initial data $\l_0\equiv\a_0$ is small and $<0$ the flow converges to a
fixed point $\Bell^*$
\be
\ell^*_0=-x_0\frac{1+5x_0}{1-4x_0},\quad
\ell^*_1=\frac{x_0}3,\quad
\ell^*_2=\frac{1}3,\quad
\ell^*_3=\frac{x_0}{18}
\Eq{e5.15}\ee
where $x_0\approx0.15878626704216...$ is the real root
of $4-19x-22x^2-107x^3=0$. The corresponding $\Ba^*$ is
(see Eq.\equ{e5.11})
\be
\a^*_0=\ell^*_0,\quad
\a^*_1=\ell^*_1,\quad \a^*_2=\ell^*_2,\quad
\a^*_3=\ell^*_3-\frac1{12}\ell^{*2}_0-\frac12\ell^*_1\ell^*_2 =  {-\frac1{12}\ell^{*2}_0}
\Eq{e5.16}\ee

\*

\0{\it Remark}:
Proving that the flow converges to $\Bell^*$ analytically is complicated by the
somewhat contrived expression of $\Bell^*$.
It is however not difficult to prove
that if the flow converges, then it must go to $\Bell^*$ (see App.\ref{appE}).
Since the numerical iterations
of the flow converge quite clearly, we will not attempt a full proof of the
convergence to the fixed point.
\*

\0{\it Remark}: A simpler case that can be treated analytically is that in which
the {\it irrelevant} terms ($\ell_1$ and $\ell_3$) are neglected (the flow in this
case is (at least numerically) {\it close} to that of the full beta function in
Eq.(\ref{e5.13}) projected onto $\ell_1=\ell_3=cst$).
Indeed the map reduces to
\begin{eqnarray}
C^{[m]}&=&1+
3\ell_0^2+9\ell_2^2\nn\\
\ell_0^{[m-1]}&=&\frac1{C^{[m]}}\Big(\ell_0
+3 \ell_0\ell_2
-2\ell_0^2\Big)
\Eq{e5.17}\\
\ell_2^{[m-1]}&=&
\frac1{C^{[m]}}\Big(2\ell_2+ \ell_0^2\Big)\nn\\
\nn\end{eqnarray}
which can be shown to have $4$ fixed points:

\0$f_0=(0,0)$, unstable in the $\ell_2$ direction and marginal in the
$\ell_0$ direction (repelling if $\ell_0<0, \ell_2=0$), this is the
{\it trivial fixed point};\\
$f_+=(0,\frac13)$, stable in the $\ell_2$ direction and marginal in
the $\ell_0$ direction (repelling if $\ell_0<0, \ell_2=\frac13$),
which we call the {\it ferromagnetic fixed-point} (because the flow
converges to $f_+$ in the ferromagnetic case, see below);\\
$f_-=(0,-\frac13)$ stable in both directions;\\
$f^*=(-\frac23,\frac13)$, stable in both directions, which we call the
{\it anti-ferromagnetic fixed point} (because the flow converges to
$f^*$ in the anti-ferromagnetic case, see below).

One can see by straightforward computations that the flow starting from
$-\frac23<\ell^{[0]}_0<0$ and $\ell^{[0]}_2=0$ converges to $f^*$ and that the
flow starting from $\ell^{[0]}_0>0$ and $\ell^{[0]}_2=0$ converges to
$f_+$ (see App.\ref{appE}).

%%%%%%%%%%%%%%%%%%%%%%%%%%%%%%%%%%%%%%%%%%%%%%%
%%%%%%%%%%%%%%%%%%%%%%%%%%%%%%%%%%%%%%%%%%%%%%%
\section{Beta function for the Kondo effect}
\label{sec6}
%%%%%%%%%%%%%%%%%%%%%%%%%%%
%%%%%%%%%%%%%%%%%%%%%%%%%%%

In this section, we discuss the Kondo effect in the hierarchical model: \ie
the phenomenon that as soon as the interaction is strictly repulsive (\ie
$\l_0<0$) the susceptibility of the impurity at zero temperature remains
positive and finite, although it can become very large for small coupling.
The problem will be rigorously reduced to the
study of a dynamical system, extending the map $\Bell\to\RR\Bell$ in
Eq.\equ{e5.13}. The value of the susceptibility follows from the iterates of
the map, as explained below. The computation will be
performed numerically; a rigorous computer assisted analysis of the flow
appears possible, but we have not attempted it because the results are very
stable and clear.
\*

We introduce a magnetic field of amplitude $h\in R$ and direction
$\Bo\in\mathcal S_2$ (in which $\mathcal S_2$ denotes the $2$-sphere)
acting on the impurity. As a consequence, the potential $V$ becomes
\be
V(\ps,\f)=V_0(\ps,\f)
-h\sum_{j\in\{1,2,3\}}
\int dt(\f_{\a}^+(t)\sigma_{\a,\a'}^j\f_{\a'}^-(t))\, \o_j
\Eq{e6.1}\ee
The corresponding partition function is denoted by
$Z_h\defi\media{e^{-V_h}}$ and the free energy of the system by
$f_h\defi-\beta^{-1}\log Z_h$. The {\it impurity susceptibility} is then
defined as
\be
\chi(h,\b)\defi \frac{\partial^2f_h}{\partial h^2}.
\Eq{e6.2}\ee
The $h$-dependent potential and the constant term, \ie $-V^{[m]}_h$ and
$c^{[m]}_h$, are then defined in the same way as in Eq.\equ{e5.4}, in terms
of which,
\be
f_h=\sum_{m=-N_\b}^0c^{[m]}_h.
\Eq{e6.3}\ee
\*
We compute $c^{[m]}_h$ in the same way as in Sec.\ref{sec5}. Because of
the extra term in the potential in Eq.\equ{e6.1}, the number of running
coupling constants increases to nine: indeed we prove by induction that
$V^{[m]}_h$ is parametrized by nine real numbers,
$\Ba^{[m]}_h=(\a_{0,h}^{[m]},\cdots,\a_{8,h}^{[m]})\in R^9$:
\be
-V_h^{[m]} (\ps^{[\le m]},\f^{[\le m]})
=\sum_{\Delta\in\mathcal Q_m}
\sum_{n=0}^8\alpha_{n,h}^{[m]}\sum_{\eta\in\{-,+\}}O_{n,\eta}^{[\le m]}(\D)
\Eq{e6.4}\ee
where $O_{n,\eta}^{[\le m]}(\D)$ for $n\in\{0,1,2,3\}$
was defined in Eq.\equ{e5.6} and
\begin{eqnarray}
O^{[\le m]}_{4,\eta}(\D)&\defi&\frac12\V A^{[\le m]}_\eta(\D)\cdot\Bo\nn\\
O^{[\le m]}_{5,\eta}(\D)&\defi&\frac12\V B^{[\le m]}_\eta(\D)\cdot\Bo\nn\\
O^{[\le m]}_{6,\eta}(\D)&\defi&\frac12\Big(\V A^{[\le m]}_\eta(\D)
\cdot\Bo\Big)\Big(\V B^{[\le m]}_\eta(\D)\cdot\Bo\Big)\Eq{e6.5}\\
O^{[\le m]}_{7,\eta}(\D)&\defi&
\frac12\Big(\V A^{[\le m]}_\eta(\D)
\cdot\V A^{[\le m]}_\eta(\D)\Big)
\Big(\V B^{[\le m]}_\eta(\D)
\cdot\Bo\Big)
\nn\\
O^{[\le m]}_{8,\eta}(\D)&\defi&\frac12\Big(\V B^{[\le m]}_\eta(\D)
\cdot\V B^{[\le m]}_\eta(\D)\Big)
\Big(\V A^{[\le m]}_\eta(\D)
\cdot\Bo\Big).\nn
\end{eqnarray}
We proceed as in Sec.\ref{sec5}. For $m=0$,
we write $V_h(\ps,\f)$ as in Eq.\equ{e6.4}
with $\Ba_h=(1,0,0,0,0,h,0,0,0)$. For $m<0$, we rewrite
\be
\big\langle\exp{\sum_{\D}}\sum_{n,\eta}\alpha_{n,h}^{[m]}
O_{n,\eta}^{[\le m]}(\D)\big\rangle_m
=\prod_{\D}\Big\langle\prod_{\eta=\pm}
\sum_{k=0}^4\frac1{k!}\Big({\sum_{n=0}^8}
\alpha_{n,h}^{[m]}O_{n,\eta}^{[\le m]}(\D)\Big)^k\Big\rangle_m
\Eq{e6.6}\ee
and, using lemma~\ref{Olemma}, we rewrite
\be
\sum_{k=0}^4\frac1{k!}\Big({\sum_{n=0}^8} \alpha_{n,h}^{[m]}
O_{n,\eta}^{[\le m]}(\D)\Big)^k
=1+\sum_{n=0}^8\ell_{n,h}^{[m]}
O_{n,\eta}^{[\le m]}(\D)
\Eq{e6.7}\ee
where $\ell_{n,h}^{[m]}$ is related to $\a_{n,h}^{[m]}$ by Eq.\equ{eC.2}.
Inserting Eq.\equ{e6.7} into Eq.\equ{e6.6} the average is
evaluated, although the
computation is even longer than that in Sec.\ref{sec5}, but can be
performed easily using a computer (see App.\ref{appI}). The result of the computation is a map
$\wt{\mathcal R}$ which maps $\ell_{n,h}^{[m]}$ to $\ell_{n,h}^{[m-1]}$, as
well as the expression for the constant $C_h^{[m]}$. Their explicit
expression is somewhat long, and is deferred to Eq.\equ{eC.1}.

By Eq.\equ{e5.14}, we rewrite Eq.\equ{e6.2} as
\be
\chi(h,\b)=\sum_{m=-N_\b}^02^m\Big(\frac{\partial_h^2
C_h^{[m]}}{C_h^{[m]}}-\frac{(\partial_h C_h^{[m]})^2}{(C_h^{[m]})^2}\Big).
\Eq{e6.8}\ee
In addition, the derivatives of $C_h^{[m]}$ can be computed exactly using
the flow in Eq.\equ{eC.1}: indeed $\partial_hC_h^{[m]}=\partial_\Bell
C_h^{[m]}\cdot\partial_h\Bell_h^{[m]}$ and similarly for
$\partial^2_hC_h^{[m]}$, and $\partial_h\Bell_h^{[m]}$ can be computed
inductively by deriving
$\wt{\mathcal R}(\Bell)$:
\be
\partial_h\Bell_h^{[m-1]}=
\partial_\Bell\wt{\mathcal R}(\Bell_h^{[m]}) \cdot
\partial_h\Bell_h^{[m]},
\Eq{e6.9}\ee
and similarly for $\partial_h^2\Bell_h^{[m]}$.
Therefore, using Eq.\equ{eC.1} and its derivatives, we can
inductively compute $\chi(\b,h)$.
\*

\begin{figure}
\hfil\includegraphics[width=200pt]{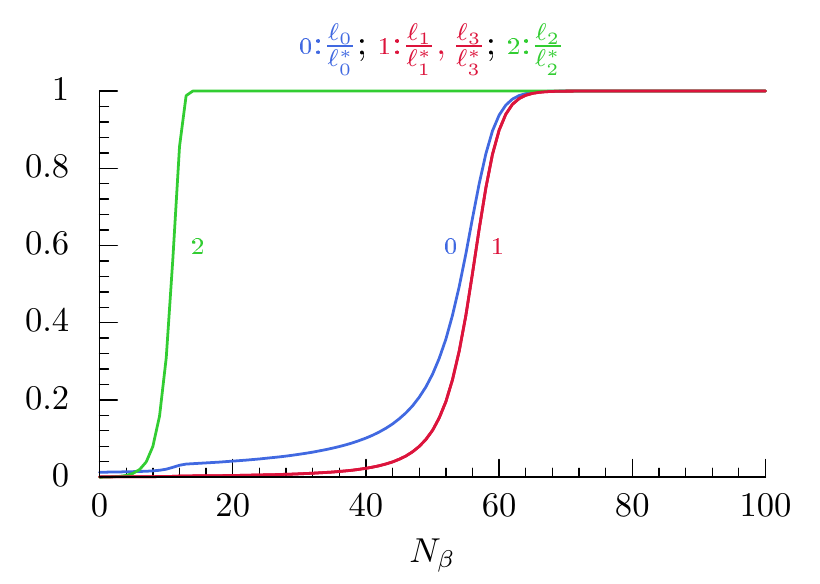}
\caption{
  plot of $\frac{\Bell}{\Bell^*}$ as a function of the iteration step $N_\b$
  for $\l_0\equiv\alpha_0=-0.01$. The {\it relevant} coupling $\ell_2$
  (curve number 2, in
  {\color{igreen}green}, color online) reaches its fixed point first, after
  which the {\it marginal} coupling $\ell_0$ (number 0, {\color{iblue}blue}) tends to
  its fixed value, closely followed by the irrelevant couplings $\ell_1$ and
  $\ell_3$ (number 1, both are drawn in {\color{ired}red} since they are almost
  equal).}
\label{fig1}
\end{figure}

By a numerical study which produces results that are
stable and clear we find that:
\*

\0(1) if $\l_0\equiv \a_0<0$, $\a_j=0,\, j>0$, and $h=0$, then the flow tends
to a nontrivial, $\l$--independent, fixed point $\Bell^*$ (see Fig.\ref{fig1}).

\*

\begin{figure}
\hfil\includegraphics[width=200pt]{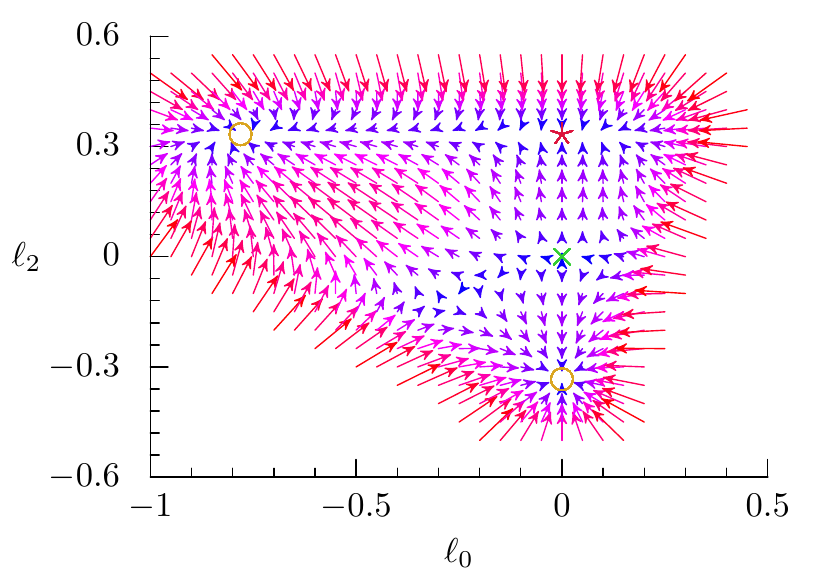}
\caption{phase diagram of the flow projected on the $(\ell_0,\ell_2)$ plane,
  with initial conditions chosen in the plane that contains all four fixed
  points: $\Bell^*$ (which is linearly stable and represented by a yellow
  circle), $\Bell^*_0$ (which has one linearly unstable direction and one
  quadratically marginal and is represented by a green cross),
  $\Bell_+^*$ (which has one
  linearly stable direction and one quadratically {marginal} and is represented by a red
  star), and $\Bell_-^*$ (which is linearly stable, and is represented by a
  yellow circle).}
\label{fig2}
\end{figure}

We define $n_j(\l_0)$ for $j=0,1,2,3$ as the step of the flow at which the
right-discrete derivative of $\ell_j/\ell_j^*$ with respect to the step $N_\b$
is largest. The reason for this definition is that, as $\l_0$ tends to $0$, the flow of $\ell_j$ tends to a
step function, so that for each component $j$ the scale $n_j$ is a good measure
of the number of iterations needed for that component to reach its fixed
value. The {\it Kondo temperature} $\b_K$ is defined as $2^{n_0(\l_0)}$, and
is the temperature at which the non-trivial fixed point is reached by all
components. For small $\l_0$, we find that (see Fig.\ref{fig7}), for $j=0,1,3$,
\be
n_j(\l_0)=c_0|\l_0|^{-1}+O(1),\quad c_0\approx0.5
\Eq{e6.10}\ee
and (see Fig.\ref{fig8})
\be
n_2(\l_0)=c_2|\log_2|\l_0||+O(1),\quad c_2\approx2.
\Eq{e6.11}\ee
\*

\0(2) In addition to the previously mentioned fixed point $\Bell^*$, there are at
least three extra fixed points, located at $\Bell_0^*\defi(0,0,0,0)$ and
$\Bell_\pm^*\defi(0,0,\pm1/3,0)$ (see Fig.\ref{fig2}).

\*

When the running coupling
constants are at $\Bell^*$, the susceptibility remains
finite as $\b\to\infty$ and positive, whereas when they are at $\Bell_+^*$, it grows
linearly with $\b$ (which is why $\Bell_+^*$ was called ``trivial'' in the
introduction).

\begin{figure}
\hfil\includegraphics[width=200pt]{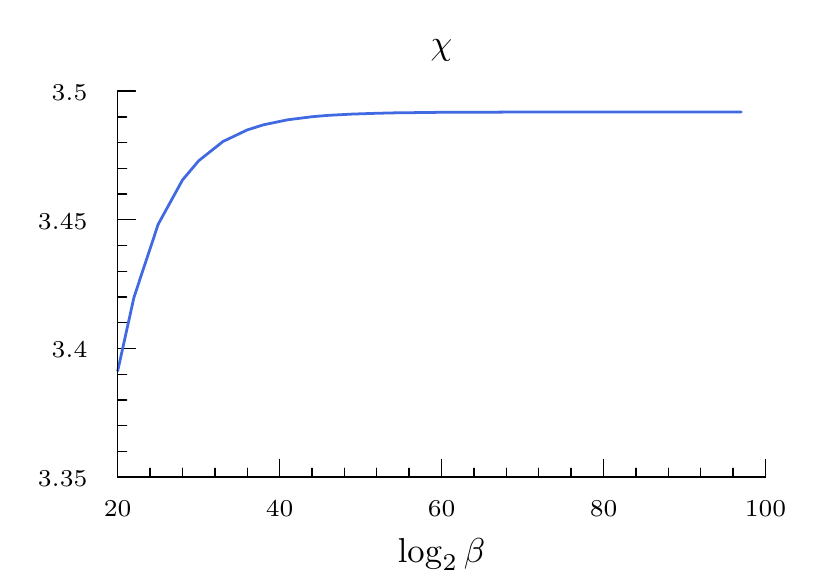}
\caption{plot of $\chi(\beta,0)$ as a function of $\log_2\beta$ for $\l_0=-0.28$.}
\label{fig3}
\end{figure}

In addition, when $\l_0<0$ the flow escapes
along the unstable direction towards the neighborhood of $\Bell^*_+$,
which is reached after $n_2(\l_0)$ steps, but since it is marginally
unstable for $\l_0<0$, it flows away towards $\Bell^*$ after $n_K(\l_0)$
steps. The susceptibility is therefore finite for $\l_0<0$ (see Fig.\ref{fig3}
(which may be compared to the exact solution \cite[Fig.3]{AFL983})).

If $\l_0>0$,
then the flow approaches $\Bell^*_+$ from the $\l_0>0$ side, which is
marginally stable, so the flow never leaves the vicinity of $\ell_+^*$
and the susceptibility diverges as $\beta\to\infty$.

\begin{figure}
\hfil\includegraphics[width=220pt]{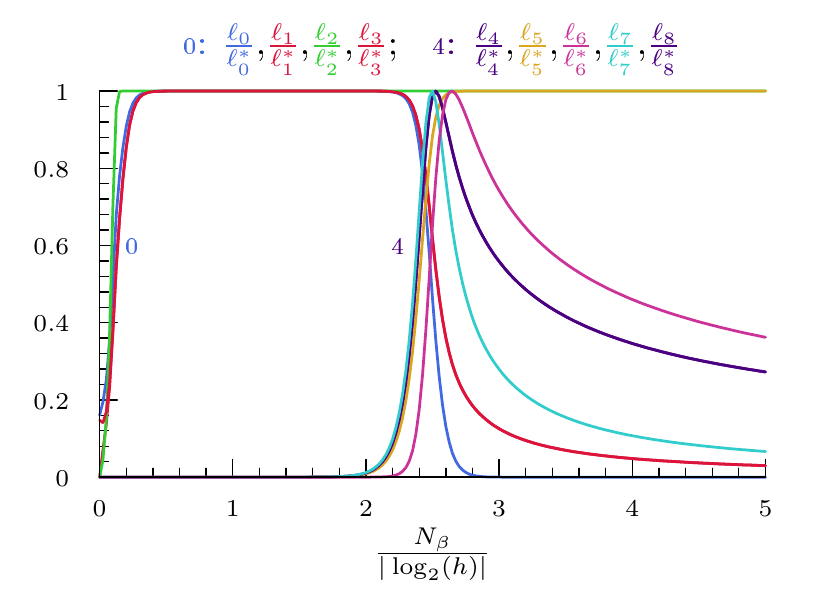}
\caption{plot of $\frac{\Bell}{\Bell^*}$ as a function of the iteration step $N_\b$ for
$\l_0=-0.01$ and $h=2^{-40}$. Here $\ell_0^*$ through $\ell_3^*$ are the
components of the non-trivial fixed point $\Bell^*$ and $\ell_4^*$ through
$\ell_8^*$ are the values reached by $\ell_4$ through $\ell_8$ of largest
absolute value. The flow behaves similarly to that at $h=0$ until $\ell_4$
through $\ell_8$ become large, at which point the couplings decay to 0,
except for $\ell_5$ and $\ell_2$.}
\label{fig4}
\end{figure}

\* \0(3) We now discuss the flow at $h>0$ and address the question of
continuity of the susceptibility in $h$ as $h\to0$.
If $\l_0 <0$ and $\a_5=h\ll\b_K^{-1}=2^{-n_K(\l_0)}$, $\ell_0$ through
$\ell_3$ first behave similarly to the $h=0$ case and tend to the same fixed
point $\Bell^*$ {and stay there until} $\ell_4$ through $\ell_8$ become large enough, after which
the flow tends to a fixed point in which $\ell_2=1/3$, $\ell_5=2$ and $\ell_j=0$
for $j\neq2,5$ (see Fig.\ref{fig4}).

\*

Setting the initial conditions for the flow as $\a_j=\a_j^*$ for $j=0,1,2,3$
and $\a_5=h$, we define $ r_j(h)$ for $j=0,1,3$ and $j=4,5,6,7,8$ as the step of
the flow at which the discrete derivative of $\ell_j/\ell_j^*$ is respectively
smallest (that is most negative) and largest. Thus $ r_j(h)$ measures when
the flow leaves $\Bell^*$. We find that (see Fig.\ref{fig9}) for small $h$,
\be
r_j(h)=c_r\log_2 h^{-1}+O(1),\quad c_r\approx2.6.
\Eq{e6.12}\ee
\*

\0Note that the previous picture only holds if $r_j(h)\gg\log_2(\b_K)$, that
is $\b_Kh\ll1$.
\*

\begin{figure}
\hfil\includegraphics[width=200pt]{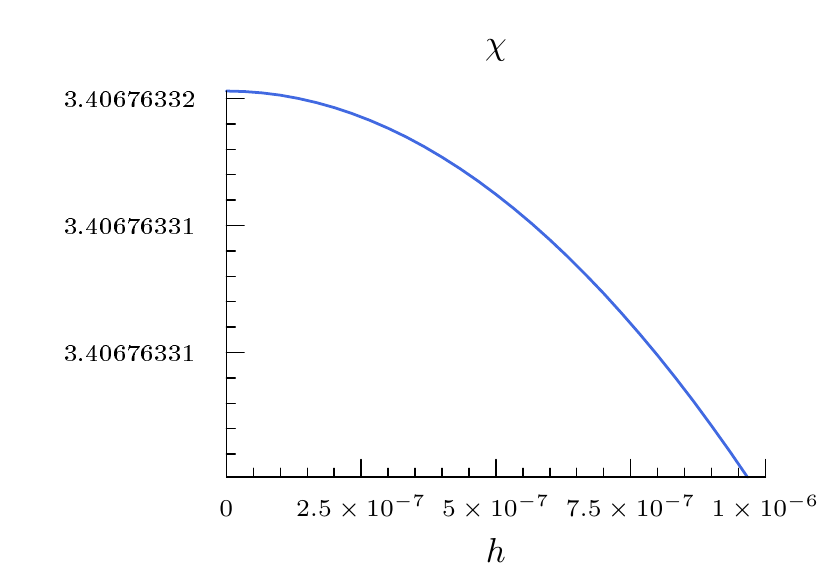}
\caption{plot of $\ch(\b,h)$ for $h\le10^{-6}$ at $\l_0=-0.28$ and
  $\b=2^{20}$ (so that the largest value for $\b h$ is $\sim1$).}
\label{fig5}
\end{figure}

The susceptibility at $0<h\ll\beta_K^{-1}$ is continuous in $h$ as
$h\to0$ (see Fig.\ref{fig5}).
This, combined with the discussion in point (2) above,
implies that the hierarchical Kondo model exhibits a Kondo effect.

\begin{figure}
\hfil\includegraphics[width=200pt]{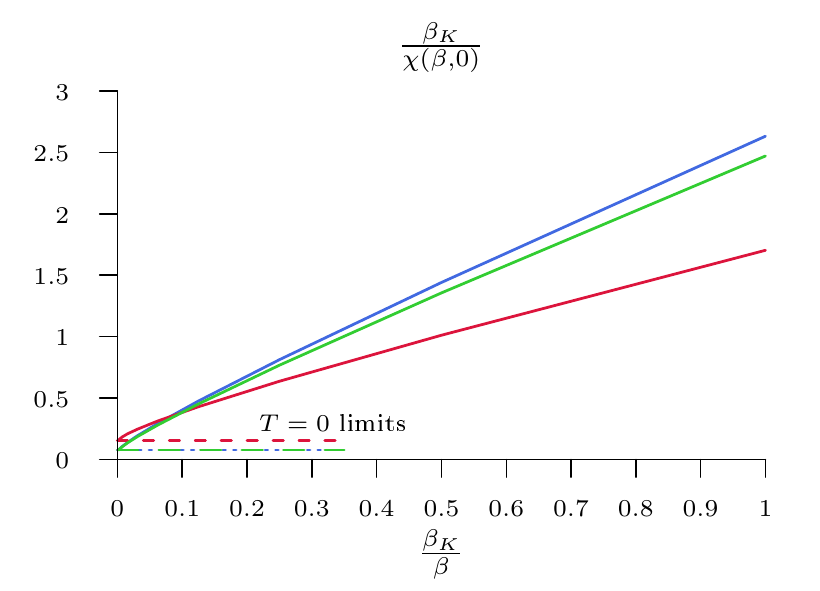}
\caption{plot of $\frac{\b_K}{\chi(\b,0)}$ as a function of
$\frac{\b_K}{\b}$ for various values of $\l_0$:
$\l_0=-0.024$ ({\color{iblue}blue}), $\l_0=-0.02412$ ({\color{igreen}green}),
$\l_0=-0.05$ ({\color{ired}red}). In \cite{Wi975}, $\l_0=-0.024$ and $-0.02412$.  Note that the abscissa of the data points are $2^{-n}$ for
$n\ge0$, so that there are only $3$ points in the range $[0.25,1]$. The
lines are drawn for visual aid.}
\label{fig6}
\end{figure}

\*

\0(4) In \cite[Fig.17, p.836]{Wi975}, there is a plot of $\frac{\b_K}{\chi(\b,0)}$ as a function of
$\frac{\b_K}\b$. For the sake of comparison, we have reproduced it for
the hierarchical Kondo model (see Fig.\ref{fig6}).

Similarly to \cite{Wi975}, we find that $\frac{\b_K}\chi$ seems
to be affine as it approaches the Kondo temperature, although it is
hard to tell for sure because of the scarcity of data points (by its
very construction, the hierarchical Kondo model only admits inverse
temperatures that are powers of 2 so the portion of Fig.\ref{fig6} that
appears to be affine actually only contains three data points).
However, we have found that such
a diagram depends on $\l_0$: indeed, by sampling values of $|\l_0|$
down to $10^{-4}$, $\frac{\b_K}{\chi(\b,0)}$ has been
found to tend to 0 faster than $(\log\b_K)^{-1.2}$ but slower than
$(\log\b_K)^{-1.3}$. In order to get a more precise estimate on this
exponent, one would need to consider $|\l_0|$ that are smaller than
$10^{-4}$, which would give rise to numerical values larger than
$10^{5000}$, and since the numbers used to perform the numerical
computations are {\it {\tt x86}-extended precision floating point
numbers}, such values are too large.

%%%%%%%%%%%%%%%%%%%%%%%%%%%%%%%%%%%%%%%%%%%%%%%%%%%%%%%
%%%%%%%%%%%%%%%%%%%%%%%%%%%%%%%%%%%%%%%%%%%%%%%%%%%%%%%
\section{Concluding remarks}
\label{sec7}
%%%%%%%%%%%%%%%%%%%%%%%%%%%%%%%%%%%%%%%%%%%%%%%%%%%%%%%
%%%%%%%%%%%%%%%%%%%%%%%%%%%%%%%%%%%%%%%%%%%%%%%%%%%%%%%

\indent(a) The hierarchical Kondo model defined in Sec.\ref{sec4} is a well
defined statistical mechanics model, for which the partition function and
correlation functions are unambiguously defined and finite as long as $\b$
is finite.  In addition, since the magnetic susceptibility of the impurity
can be rewritten as a correlation function:
\be
\chi(\b,0)=\int_0^\b dt\,
\media{(\f^+(0)\Bs\f^-(0))(\f^+(t)\Bs\f^-(t))}_{h=0},
\Eq{e7.1}\ee
$\chi(\b,0)$ is a thermodynamical quantity of the model.
\*
(b) The qualitative behavior of the renormalization group flow is unchanged
if all but the relevant and marginal running coupling constants
(\ie six constants out
of nine) of the beta functions of Sec.\ref{sec5},\ref{sec6} are neglected
(\ie set to $0$ at every step of the iteration). In particular, we still
find a Kondo effect.
\* (c) In the hierarchical model defined in Sec.\ref{sec4}, quantities
other than the magnetic susceptibility of the impurity can be computed,
although all observables must only involve fields localized at $x=0$.  For
instance, the response to a magnetic field acting on all sites of the
fermionic chain as well as the impurity cannot be investigated in this
model, since the sites of the chain with $x\neq0$ are not accounted for.

\hskip.5cm(c.a) We have attempted to extend the definition of the
hierarchical model to allow observables on the sites of the chain at $x\neq0$
by paving the space-time
plane with square boxes (instead of paving the time axis with intervals,
see Sec.\ref{sec4}), defining hierarchical fields for each quarter box and
postulating a propagator between them by analogy with the non-hierarchical
model. The magnetic susceptibility of the impurity is defined as the
response to a magnetic field acting on every site of the chain and on the
impurity, to which the susceptibility of the non-interacting chain is
subtracted. We have found, iterating the flow numerically, that for such a
model {\it there is no Kondo effect}, that is the impurity susceptibility
diverges as $\b$ when $\b\to\infty$.

\hskip.5cm(c.b) A second approach has yielded better results, although it
is not completely satisfactory. The idea is to incorporate the effect of
the magnetic field $h$ acting on the fermionic chain into the propagator of
the non-hierarchical model, after which the potential $V$ only depends on
the site at $x=0$, so that the hierarchical model can be defined in the
same way as in Sec.\ref{sec4} but with {\it an $h$-dependent
  propagator}. In this model, we have found that {\it there is a Kondo
  effect}.

%%%%%%%%%%%%%%%%%%%%%%%%%%%%%%%%%%%%%%%%%%%%%%%%%%%%%%%
%%%%%%%%%%%%%%%%%%%%%%%%%%%%%%%%%%%%%%%%%%%%%%%%%%%%%%%
\appendix

%%%%%%%%%%%%%%%%%%%%%%%%%%%%%%%%%%%%%%
%%%%%%%%%%%%%%%%%%%%%%%%%%%%%%%%%%%%%
\section{Comparison with the original Kondo model}
\label{appA}
%%%%%%%%%%%%%%%%%%%%%%%%%%%%%%%%%%%%%%
%%%%%%%%%%%%%%%%%%%%%%%%%%%%%%%%%%%%%%

If the partition function for the original Kondo model in presence of a
magnetic field $h$ acting only on the impurity site and at finite $L$
is denoted by $Z^0_K(\b,\l_0,h)$
and the partition function for the model Eq.\equ{e2.1} with the same field $h$ is denoted by
$Z_K(\b,\l_0,h)$, then
\be
Z_K(\b,\l_0,h)=Z^0_K(\b,\l_0,h)+Z^0_K(\b,0,0)
\Eq{eA.1}\ee
so that by defining
\begin{equation}
\k\,\defi\,1+\frac{Z^0_K(\b,0,0)}{Z^0_K(\b,\l_0,h)}
\Eq{eA.2}\end{equation}
we get
\begin{eqnarray}
m_K(\b,\l_0,h)&=&\frac1\k{m^0_K(\b,\l_0,h)}, \nn\\ m^0_K(\b,\l_0,h)&=&\k\,
m_K(\b,\l_0,h)
\Eq{eA.3}\\
\ch_K(\b,\l_0,h)&=&\frac1\k\,{\ch_K^0(\b,\l_0,h)}
+\frac{\k-1}\k\,\b\,m^0_K(\b,\l_0,h)^2\nn\\
\ch_K^0(\b,\l_0,h)&=&\k\, \ch_K(\b,\l_0,h)-(\k-1)\b m_K(\b,\l_0,h)^2.\nn
\end{eqnarray}
In addition $1\le\k\le2$: indeed the first inequality is trivial
and the second follows from the variational principle
(see \cite[theorem 7.4.1, p.188]{Ru969}):
\begin{eqnarray}
\log Z^0_K(\b,\l_0,h)&=&\max_\m(s(\m)-\m(H_0+V))\nn\\
&\ge&
s(\m_0)-\m_0(H_0)+\m_0(V)=s(\m_0)-\m_0(H_0)
=\log Z^0_K(\b,0,0)
\Eq{eA.4}\end{eqnarray}
where $s(\m)$ is the entropy of the state $\m$, and in which we used
\be
\m_0(V)={\rm Tr}\,( e^{-\b H_0}\,V)/Z_K(\b,0,0)=0.
\Eq{eA.5}\ee

Therefore, for $\beta h^2\ll1$ (which implies that if there is a Kondo
effect then $\beta m_K^2\ll1$), the model Eq.\equ{e2.1} exhibits a
Kondo effect if and only if the original Kondo model does, therefore,
for the purposes of this paper, both models are {\it equivalent}.

%%%%%%%%%%%%%%%%%%%%%%%%%%%%%%%%%%%%%%
%%%%%%%%%%%%%%%%%%%%%%%%%%%%%%%%%%%%%%
\section{Some identities.}
\label{appB}
%%%%%%%%%%%%%%%%%%%%%%%%%%%%%%%%%%%%%%
%%%%%%%%%%%%%%%%%%%%%%%%%%%%%%%%%%%%%%

In this appendix, we state three relations used to compute the flow equation
Eq.\equ{e5.13}, which follow from a patient algebraic meditation:
\begin{eqnarray}
\media{A_1^{j_1}A_2^{j_2}}&=&\d_{j_1,j_2}\Big(2
+\frac13\V a^2\Big)-2\,a^{j_1,j_2}\d_{j_1\ne j_2}\, s_{t_2,t_1} \nn\\
\media{A^{j_1}_1A^{j_2}_1A^{j_3}_2}
&\equiv& 2\,a^{j_3}\,\d_{j_1,j_2}\Eq{eB.1}\\
\media{A_1^{j_1}A_1^{j_2}A_2^{j_3}A_2^{j_4}}
&=&4\d_{j_1,j_2}\d_{j_3,j_4}\nn\end{eqnarray}
where the lower case $\V a$ denote $\media{\V A_1}\equiv\media{\V
A_2}$ and $a^{j_1,j_2}=\media{\ps^+_1\s^{j_1}\s^{j_2}\ps^-_1}
=\media{\ps^+_2\s^{j_1}\s^{j_2}\ps^-_2}$.
%

%%%%%%%%%%%%%%%%%%%%%%%%%%%%%%%%%%%%%%%%%%%%%%%%%%%%%%%
\section{Complete beta function}
\label{appC}
%%%%%%%%%%%%%%%%%%%%%%%%%%%%%%%%%%%%%%%%%%%%%%%%%%%%%%%
%%%%%%%%%%%%%%%%%%%%%%%%%%%%%%%%%%%%%%%%%%%%%%%%%%%%%%%
%%%%%%%%%%%%%%%%%%%%%%%%%%%%%%%%%%%%%%%%%%%%%%%%%%%%%%%

The beta function for the flow described in Sec.\ref{sec6} is
\begin{eqnarray}
\ell_0^{[m-1]}&=&\frac1{C^{[m]}}(
\ell_0
-2\ell_0\ell_6
+18\ell_0\ell_3
+3 \ell_0\ell_2
+3 \ell_0\ell_1
-2\ell_0^2)\nn\\
%%%%%%%%%%%%%%%%%%%%%%%%%
\ell_1^{[m-1]}&=&\frac1{C^{[m]}}(
\frac12\ell_1
+9\ell_2\ell_3
+\frac32\ell_8^2
+\frac1{12}\ell_6^2
+\frac12\ell_5\ell_7
+\frac1{24}\ell_4^2
+\frac16\ell_0\ell_6
+\frac14\ell_0^2)
\nn\\
%%%%%%%%%%%%%%%%%%%%%%%%%
\ell_2^{[m-1]}&=&\frac1{C^{[m]}}(
2\ell_2+36\ell_1\ell_3
+ \ell_0^2
+6\ell_7^2
+\frac13\ell_6^2
+\frac1{6}\ell_5^2
+2\ell_4\ell_8
+\frac23\ell_0\ell_6)
\Eq{eC.1}\\
%%%%%%%%%%%%%%%%%%%%%%%%%
\ell_3^{[m-1]}&=&\frac1{C^{[m]}}(
\frac12\ell_3
+\frac14\ell_1\ell_2
+\frac1{24} \ell_0^2
+\fra1{36}\ell_0\ell_6
+\frac1{72}\ell_6^2
+\frac1{12}\ell_5\ell_7
+ \frac1{12}\ell_4\ell_8
)\nn\\
%%%%%%%%%%%%%%%%%%%%%%%%%
\ell_4^{[m-1]}&=&\frac1{C^{[m]}}(
\ell_4
+6\ell_6\ell_7
+\ell_5\ell_6
+108\ell_3\ell_8
+18\ell_2\ell_8
+3\ell_1\ell_4
+6\ell_0\ell_7
+\ell_0\ell_5
)\nn\\
%%%%%%%%%%%%%%%%%%%%%%%%%
\ell_5^{[m-1]}&=&\frac1{C^{[m]}}(
2\ell_5
+12 \ell_6\ell_8
+2\ell_4\ell_6
+216 \ell_3\ell_7
+6\ell_2\ell_5
+36\ell_1\ell_7
+12\ell_0\ell_8
+2\ell_0\ell_4)\nn\\
%%%%%%%%%%%%%%%%%%%%%%%%%
\ell_6^{[m-1]}&=&\frac1{C^{[m]}}(
\ell_6
+18\ell_7\ell_8
+3\ell_5\ell_8
+3\ell_4\ell_7
+\frac12\ell_4\ell_5
+18\ell_3\ell_6
+3\ell_2\ell_6
+3\ell_1\ell_6\nn\\
&&\hskip25pt+2\ell_0\ell_6)\nn\\
%%%%%%%%%%%%%%%%%%%%%%%%%
\ell_7^{[m-1]}&=&\frac1{C^{[m]}}(
\frac12\ell_7
+\frac12\ell_6\ell_8
+\frac1{12}\ell_4\ell_6
+\frac32\ell_3\ell_5
+\frac32\ell_2\ell_7
+\frac14\ell_1\ell_5
+\frac12\ell_0\ell_8\nn\\
&&\hskip25pt+\frac1{12}\ell_0\ell_4)\nn\\
%%%%%%%%%%%%%%%%%%%%%%%%%
\ell_8^{[m-1]}&=&\frac1{C^{[m]}}(
\ell_8
+\ell_6\ell_7
+\frac1{6}\ell_5\ell_6
+3\ell_3\ell_4
+\frac12\ell_2\ell_4
+3\ell_1\ell_8
+\ell_0\ell_7
+\frac1{6}\ell_0\ell_5)\nn\\
%%%%%%%%%%%%%%%%%%%%%%%%%
C^{[m]}&=&1+
2\ell_0^2+(\ell_0+\ell_6)^2
+9\ell_1^2
+9\ell_2^2
+324\ell_3^2
+\frac12\ell_4^2
+\frac12\ell_5^2
+18\ell_7^2
+18\ell_8^2
\nn
\end{eqnarray}
in which we dropped the $^{[m]}$ exponent on the right side.  By
considering the linearized flow equation (around $\ell_j=0$), we find that
$\ell_0,\ell_4,\ell_6,\ell_8$ are {\it marginal}, $\ell_2,\ell_5$ {\it relevant}
and $\ell_1,\ell_3,\ell_7$ {\it irrelevant}. The consequent linear flow is
{\it very different} from the full flow discussed in Sec.\ref{sec6}.

The vector $\Bell$ is related to $\Ba$ and via the following map:
\begin{eqnarray}
\ell_0&=&\a_0,\quad \ell_1=\a_1+\frac1{12}\a_4^2,\quad
\ell_2=\a_2+\frac1{12}\a_5^2\nn\\
\ell_3&=&\a_3+\frac1{12}\a_0^2+\frac1{18}\a_0\a_6+\frac12\a_1\a_2
+\frac1{6}\a_4\a_8
+\frac1{6}\a_5\a_7+\frac1{36}\a_6^2
\nn\\&&
+\frac1{36}\a_0\a_4\a_5
+\frac1{24}\a_1\a_5^2
+\frac1{24}\a_2\a_4^2%\nn\\&&
+\frac1{36}\a_4\a_5\a_6
+\frac1{288}\a_4^2\a_5^2
\Eq{eC.2}\\
\ell_4&=&\a_4,\quad
\ell_5=\a_5,\quad
\ell_6=\a_6+\frac12\a_4\a_5\nn\\
\ell_7&=&\a_7+\frac1{6}\a_0\a_4+\frac12\a_1\a_5+\frac1{6}\a_4\a_6
+\frac1{24}\a_4^2\a_5
\nn\\
\ell_8&=&\a_8+\frac1{6}\a_0\a_5+\frac12\a_2\a_4+\frac1{6}\a_5\a_6
+\frac1{24}\a_4\a_5^2.\nn\end{eqnarray}
%

%%%%%%%%%%%%%%%%%%%%%%%%%%%%%%%%%%%%%%
%%%%%%%%%%%%%%%%%%%%%%%%%%%%%%%%%%%%%%
\section{The algebra of the  operators \texorpdfstring{$O_{n,\pm}$}.. }
\label{appD}
%%%%%%%%%%%%%%%%%%%%%%%%%%%%%%%%%%%%%%
%%%%%%%%%%%%%%%%%%%%%%%%%%%%%%%%%%%%%%

\begin{lemma}
\label{Olemma}
Given $\eta\in\{-,+\}$, $m\le0$ and $\D\in\mathcal Q_m$, the
span of the operators $\{O_{n,\eta}^{[\le m]}(\D)\}_{n\in\{0,1,2,3\}}$
defined in Eq.\equ{e5.6} is an algebra, that is all linear combinations of
products of $O_{n,\eta}^{[\le m]}(\D)$'s is itself a linear combination of
$O_{n,\eta}^{[\le m]}(\D)$'s.
\\
The same result holds for the span of the operators $\{O_{n,\eta}^{[\le
    m]}(\D)\}_{n\in\{0,\cdots,8\}}$ defined in Eq.\equ{e6.5}.
\end{lemma}

\0{\it Proof}: The only non-trivial part of this proof is to show that
the product of two $O_{n,\eta}$'s is a linear combination of
$O_{n,\eta}$'s.

Due to the anti-commutation of Grassmann variables, any linear combination
of $\ps_{\a}^{[\le m]\pm}$ and $\f_{\a}^{[\le m]\pm}$ squares to
0. Therefore, a straightforward computation shows that
$\forall(i,j)\in\{1,2,3\}^2$,
\be
A^{i}_\eta A^{j}_\eta=2\delta_{i,j}
\ps^{+}_{\uparrow}
\ps^{+}_{\downarrow}
\ps^{-}_{\uparrow}
\ps^{-}_{\downarrow},\quad
B^{i}_\eta B^{j}_\eta=2\delta_{i,j}
\f^{+}_{\uparrow}
\f^{+}_{\downarrow}
\f^{-}_{\uparrow}
\f^{-}_{\downarrow}
\Eq{eD.1}\ee
where the labels $^{[\le m]}$ and $(\D)$ are dropped to alleviate the
notation. In particular, this implies that any product of three $A^{i}_\eta$
for $i\in\{1,2,3\}$ vanishes (because the product of the
right side of the first of Eq.\equ{eD.1} and any Grassmann field
$\ps_{\a}^{\pm}$ vanishes) and similarly for the product of three
$B^{i}_\eta$.

Using Eq.\equ{eD.1}, we prove that $\mathrm{span}\{O_{n,\eta}^{[\le
    m]}(\D)\}_{n\in\{0,1,2,3\}}$ is an algebra. For all $n\in\{0,1,2,3\}$,
$p\in\{1,2,3\}$, $l\in\{1,2\}$,
\be
O_{p}^2=0,\quad
O_{3}O_{n}=0,\quad
O_{l}O_{0}=0,\quad
O_{0}^2=\frac16O_{3},\quad
O_{1}O_{2}=\frac12O_{3}
\Eq{eD.2}\ee
(here the $^{[\le m]}$, $(\D)$ and $_\eta$ are dropped).
This concludes the proof of the first claim.

Next we prove that $\mathrm{span}\{O_{n,\eta}^{[\le
    m]}(\D)\}_{n\in\{0,\cdots,8\}}$ is an algebra. In addition to
Eq.\equ{eD.2}, we have, for all $p\in\{0,\cdots,8\}$,
\begin{eqnarray}
&&O_0O_4=\frac16O_7,\quad
O_0O_5=\frac16O_8,\quad
O_0O_6=\frac1{18}O_3,\quad
O_0O_7=O_0O_8=0,\quad
O_1O_5=\frac12O_7,
\nn\\&&
O_1O_4=O_1O_6=O_1O_7=O_1O_8=0,\quad
O_2O_4=\frac12O_8,\quad
O_2O_5=O_2O_6=O_2O_7=O_2O_8=0,
\nn\\&&
O_3O_p=0,\quad
O_4^2=\frac16O_1,\quad
O_4O_5=\frac12O_6,\quad
O_4O_8=\frac16O_3,\quad
O_4O_7=0,\quad
O_5^2=\frac16O_2,
\Eq{eD.3}\\&&
O_5O_7=\frac16O_3,\quad
O_5O_8=0,\quad
O_6^2=\frac1{18}O_3,\quad
O_6O_7=O_6O_8=0,\quad
O_7^2=O_8^2=O_7O_8=0.
\nn\end{eqnarray}
This concludes the proof of the lemma.

%%%%%%%%%%%%%%%%%%%%%%%%%%%%%%%%%%%%%%
%%%%%%%%%%%%%%%%%%%%%%%%%%%%%%%%%%%%%
\section{Fixed points at $h=0$}
\label{appE}
%%%%%%%%%%%%%%%%%%%%%%%%%%%%%%%%%%%%%%
%%%%%%%%%%%%%%%%%%%%%%%%%%%%%%%%%%%%%%

We first compute the fixed points of Eq.\equ{e5.13} for $\ell_2\ge0$.
It follows from Eq.\equ{e5.13} that if $\Bell$ is a fixed point, then
$\ell_1=6\ell_3$, which implies
\be
(1-3\ell_2)\left(
\ell_2(1+3\ell_2)+6\ell_1^2+\ell_0^2
\right)=0.
\Eq{eE.1}\ee
If $\ell_2\ge0$, Eq.\equ{eE.1} implies that either $\ell_2=\ell_1=\ell_0=0$
or $\ell_2=\frac13$. In the latter case, 
either $\ell_0=\ell_1=0$ or $\ell_0\not=0$ and Eq.\equ{e5.13} becomes
\be
\left\{\begin{array}l
3\ell_0^2+2\ell_0+6\ell_1(3\ell_1-1)=0\\[0.3cm]
\ell_1(1+18\ell_1^2)+\ell_0^2(3\ell_1-\frac14)=0.
\end{array}\right.
\Eq{eE.2}\ee
In particular, $\ell_1(1-12\ell_1)>0$, so that
\be
\ell_0=\pm2\sqrt{\frac{\ell_1(1+18\ell_1^2)}{1-12\ell_1}}
\Eq{eE.3}\ee
which we inject into Eq.\equ{eE.2} to find that $\ell_0<0$ and
\be
1-\frac{35}4(3\ell_1)+\frac{27}2(3\ell_1)^2-\frac{19}4(3\ell_1)^3+107(3\ell_1)^4=0.
\Eq{eE.4}\ee
Finally, we notice that $\frac1{12}$ is a solution of Eq.\equ{eE.4}, which implies
that
\be
4-19(3\ell_1)-22(3\ell_1)^2-107(3\ell_1)^3
\Eq{eE.5}\ee
which has a unique real solution.
Finally, we find that if $\ell_1$ satisfies Eq.\equ{eE.5}, then
\be
2\sqrt{\frac{\ell_1(1+18\ell_1^2)}{1-12\ell_1}}=3\ell_1\frac{1+15\ell_1}{1-12\ell_1}.
\Eq{eE.6}\ee
We have therefore shown that, if $\ell_2\ge0$, then Eq.\equ{e5.13} has three fixed
points:
\begin{eqnarray}
&&\Bell_0^*:=(0,0,0,0),\quad
\Bell_+^*:=\left(0,0,\frac13,0\right),\nn\\
&&\Bell^*:=\left(-x_0\frac{1+5x_0}{1-4x_0},\frac{x_0}3,\frac13,\frac{x_0}{18}\right).
\Eq{eE.7}\end{eqnarray}
In addition, it follows from Eq.\equ{e5.13} and Eq.\equ{e5.11} that, if $\lambda_0<0$, then
(recall that $\a_0^{[0]}=\l_0$ and $\a_i^{[0]}=0$, $i=1,2,3$)
\be
\ell_0^{[m]}<0,\quad
0\le\ell_2^{[m]}<\frac13,\quad
0\le\ell_1^{[m]}<6\ell_3^{[m]}<\frac1{12}
\Eq{eE.8}\ee
for all $m\le0$, which implies that the set $\{\Bell\ |\ \ell_0<0,\ \ell_2\ge0,\ \ell_1\ge0,\ \ell_3\ge0\}$
is stable under the flow.
In addition, if $\ell_0^{[m]}>-\frac23$, then
$\ell_0^{[m-1]}<\ell_0^{[m]}$, so that the flow cannot converge to
$\Bell_0^*$ or $\Bell_+^*$. Therefore if the flow converges, then it converges
to $\Bell^*$.
\*

We now study the {\it reduced} flow Eq.\equ{e5.17}, and
prove that starting from $-2/3<\ell^{[0]}_0<0$, $\ell^{[0]}_2=0$,
the flow converges to $f^*$. It follows from
Eq.\equ{e5.17} that $\ell^{[m]}_0<0$, $\ell^{[m]}_2>0$ for all $m<0$, so that
if Eq.\equ{e5.17} converges to a fixed point, then it must converge to $f^*$.
In addition, by a straightforward induction, one finds that
$\ell_2^{[m-1]}>\ell_2^{[m]}$ if $\ell_2^{[m]}<\frac13$. Furthermore,
$(2\ell_2^{[m]}+(\ell_0^{[m]})^2)\le\frac 13C^{[m]}$, which implies that $\ell_2^{[m]}\le\frac13$.
Therefore $\ell_2^{[m]}$ converges as $m\to-\infty$. In addition,
$\ell_0^{[m-1]}<\ell_0^{[m]}$ if $\ell_0^{[m]}>-\frac23$, and
$\ell_0^{[m]}>-\frac13-\ell_2^{[m]}\ge-\frac23$, so that $\ell_0^{[m]}$
converges as well as $m\to-\infty$. The flow therefore tends to $f^*$.

Finally, we prove that starting from $\ell^{[0]}_0>0,\ell^{[0]}_2=0$,
the flow converges to $f_+$. Similarly to the anti-ferromagnetic case,
$\ell^{[m]}_2>0$ for all $m<0$, $\ell_2^{[m]}\le\frac13$ and
$\ell_2^{[m-1]}>\ell_2^{[m]}$. In addition, by a simple induction,
if $\lambda_0<1$, then $\ell_0^{[m]}>0$
and $\ell_0^{[m]}+\frac13-\ell_2^{[m]}$ is
strictly decreasing and positive. In conclusion, $\ell_0^{[m]}$ and
$\ell_2^{[m]}$ converge to $f_+$.

%%%%%%%%%%%%%%%%%%%%%%%%%%%%%%%%%%%%%%
%%%%%%%%%%%%%%%%%%%%%%%%%%%%%%%%%%%%%%
\section{Asymptotic behavior of $n_j(\l_0)$ and $r_j(h)$}
\label{appF}
%%%%%%%%%%%%%%%%%%%%%%%%%%%%%%%%%%%%%%
%%%%%%%%%%%%%%%%%%%%%%%%%%%%%%%%%%%%%%

\begin{figure}[!h]
\hfil\includegraphics[width=200pt]{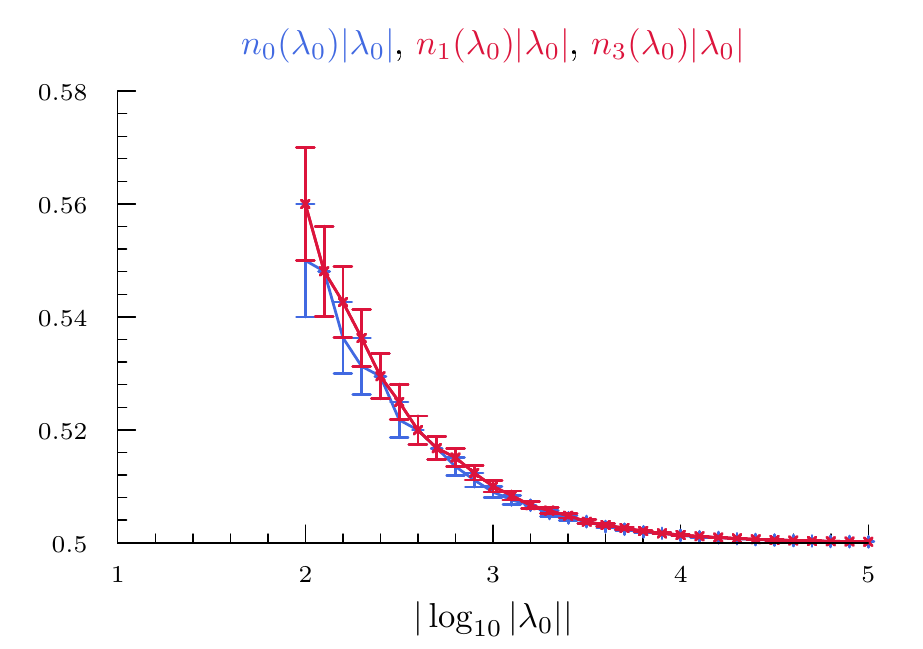}
\caption{plot of $n_j(\l_0)|\l_0|$ for $j=0$ ({\color{iblue}blue}, color
online) and $j=1,3$ ({\color{ired}red}) as a function of $|\log_{10}|\l_0||$.
This plot confirms Eq.\equ{e6.10}.}
\label{fig7}
\end{figure}

\begin{figure}[!h]
\hfil\includegraphics[width=200pt]{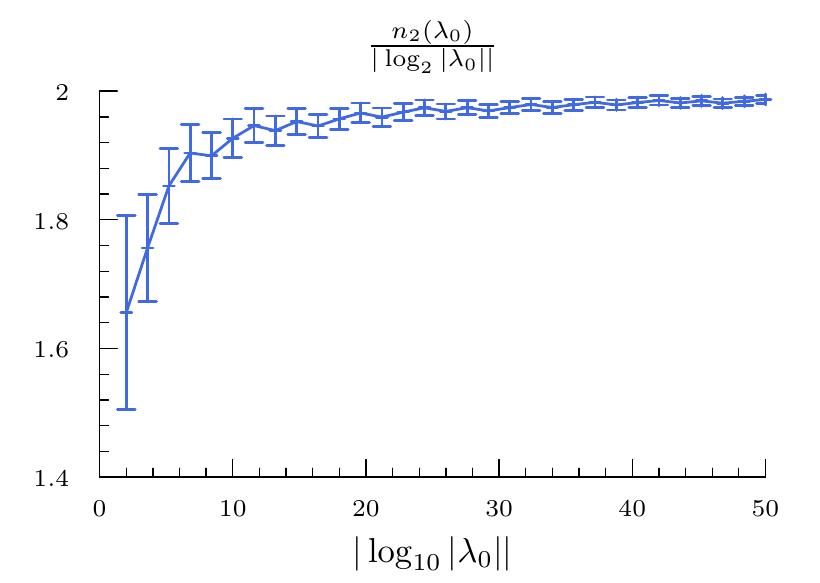}
\caption{plot of $n_2(\l_0)|\log_2|\l_0||^{-1}$ as a function of
$|\log_{10}|\l_0||$. This plot confirms Eq.\equ{e6.11}.}
\label{fig8}
\end{figure}

\begin{figure}[!h]
\hfil\includegraphics[width=200pt]{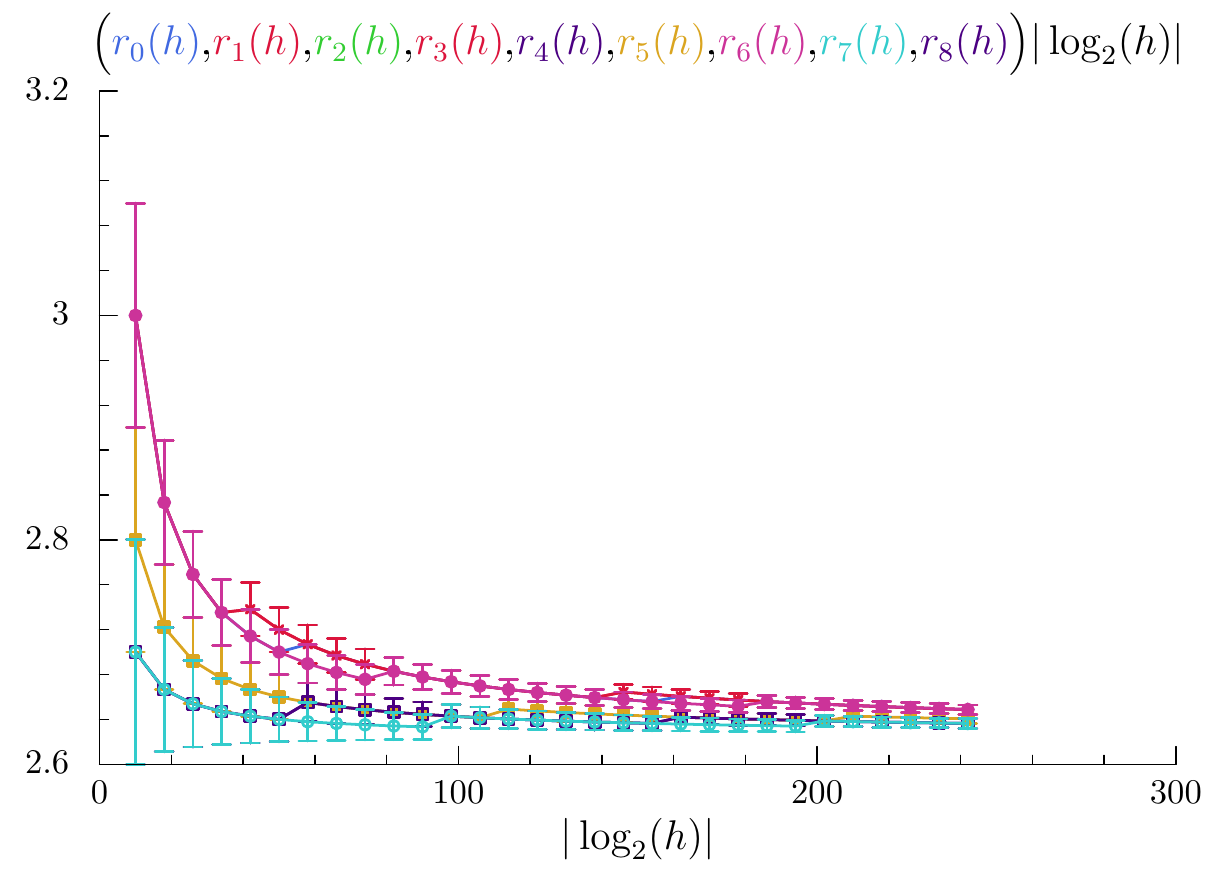}
\caption{plot of $r_j(h)|\log_2(h)||$ as a function of
$|\log_{2}(h)|$. This plot confirms Eq.\equ{e6.12}.}
\label{fig9}
\end{figure}

In this appendix, we show plots to support the claims on the asymptotic
behavior of $n_j(\l_0)$ (see Eq.\equ{e6.10}, Fig.\ref{fig7} and Eq.\equ{e6.11}, Fig.\ref{fig8}) and $r_j(h)$
(see Eq.\equ{e6.12}, Fig.\ref{fig9}). The plots below have error bars which are due to the fact
that $n_j(\l_0)$ and $r_j(h)$ are integers, so their value could be off by
$\pm1$.

%\FloatBarrier

%%%%%%%%%%%%%%%%%%%%%%%%%%%%%%%%%%%%%%
%%%%%%%%%%%%%%%%%%%%%%%%%%%%%%%%%%%%%%
\section{Kondo effect, XY-model, free fermions}
\label{appG}
%%%%%%%%%%%%%%%%%%%%%%%%%%%%%%%%%%%%%%
%%%%%%%%%%%%%%%%%%%%%%%%%%%%%%%%%%%%%%

In \cite{ABGM971}, given $\nu\in [1,\ldots,L]$, the Hamiltonian $\HH_h=\HH_0
{-h} \,\s_\nu^z$, with
\be \HH_0={- \frac 14} \sum_{n=1}^L
(\s^x_n\s^x_{n+1}+\s^y_n\s^y_{n+1})\Eq{eG.1}.
\ee
has been considered with suitable boundary conditions (see App.\ref{appH}),
under which $H_0$
and ${\s^z_0} +1$ are unitarily equivalent to $\sum_{q}{(-\cos q)} \,
a^+_qa^-_q$ and, respectively, to {$\frac2L \sum_{q,q'} a^+_q a^-_{q'}
  e^{i\nu(q-q')}$} in which $a^\pm_q$ are fermionic creation and
annihilation operators and the sums run over $q$'s that are such that
$e^{iq L}=-1$.  It has been shown, \cite{ABGM971}\footnote{\small see
  \cite{ABGM971}, Eq.(3.18) which, after integration by parts is equivalent
  to what follows. Since the scope of \cite{ABGM971} was somewhat different
  we give here a complete self-contained account of the derivation of
  Eq.\equ{eG.2} and the following ones, see App.\ref{appH}.}, that, by
defining
\begin{eqnarray} F_L(\z)&=&1+\frac{2 h}{L} \sum_q \frac1{\z+\cos q}\nn\\
 F(z)&=& \lim_{L\to\infty} F_L(z)=1+\frac{2\,h}{\p}\,\int_0^\p
\frac{dq}{(z+\cos q)}
\Eq{eG.2}\end{eqnarray}
the partition function is equal to $Z_L^0\zeta_L$ in which $Z_L^0$ is the
partition function at $h=0$ and is {\it extensive} (\ie of $O(e^{const L})$)
and {(see App.\ref{appH}, Eq.\equ{eH.12})}
\be
\log {\z_L (\b,h)}=-\b h
+\frac1{2\p i}\oint_C
\log (1+{e^{-\b z}} )
\Big[\frac{\dpr_zF_L(z)}{F_L(z)}
\Big]\,dz
\Eq{eG.3}\ee
where the contour $C$ is a closed curve which contains the zeros of
$F_L(\z)$ (\eg, {for $L\to\infty$,} a curve around the real interval
$[-1,\sqrt{1+4h^2}]$ if {$h<0$} and $[-\sqrt{1+4h^2},1]$ if {$h>0$}) but
not around those of {$1+e^{-\b\z}$} (which are on the imaginary axis and
away from $0$ by at least $\frac\p\b$). In addition, it follows from a
straightforward computation that $(F(z)-1)/h$ is equal to the analytical
continuation of {$2 (z^2-1)^{-\frac12}$} from $(1,\infty)$ to
$C\setminus[-1,1]$.

At fixed $\b<\infty$
the partition  function $\z_L(\b,h)$ has a non extensive limit $\z(\b,h)$ as
$L\to\infty$; the $\z(\b,h)$ and the
susceptibility and magnetization values $m(\b,h)$ and $\ch(\b,h)$, are given
{\it in the thermodynamic limit}
by
\begin{eqnarray}
\log \z(\b,h)&=&-\b h
{+\frac\b{2\p i}} \oint_C\frac{dz}{1+{e^{\b z}}}
\log(1 {+}  \frac{2h}{(z^2-1)^{\frac12}})\nn\\
m(\b,h)&=&-1+\frac1{\p i}\oint_C \frac1{1+{e^{\b z}}}
\frac{dz}{(z^2-1)^{\frac12}{+2h} }\Eq{eG.4}\\
\chi(\b,h)&=&-\frac{2}{\p i}\oint_C \frac1{1+{e^{\b z}}}
\frac{dz}{((z^2-1)^{\frac12}{+2h})^2}
\nn\end{eqnarray}
so that $\chi(\b,0)=\frac{2\sinh(\b)}{(1+\cosh(\b))}$ and,
in the $\b\to\infty$ limit,
\begin{equation}
m(\infty,h)=
\frac{2h}{\sqrt{1+4h^2}}
,\quad
\ch(\infty,h)=\frac2{(1+4h^2)^{3/2}}
\Eq{eG.5}\end{equation}
both of which are finite.
Adding an impurity at $0$, with spin operators $\Bt_0$, the
Hamiltonian
\be H_\l=H_0{- h}(\s^z_0+\t^z_0) {- \l}\s^z_0\t^z_0\Eq{eG.6}\ee
is obtained. Does it exhibit a Kondo effect?

Since $\Bt_0$ commutes with the $\Bs_n$ and, hence, with $H_0$, the average
magnetization and susceptibility, $m^{int}(\b,h,\l)$ and $\ch^{int}(\b,h,\l)$,
responding to a field $h$ acting only on the site $0$, can be expressed in
terms of the functions $\z(\b,h)$ and its derivatives
$\z'(\b,h)$ and $\z''(\b,h)$. By using the fact that $\z(\b,h)$ and $\z''(\b,h)$
are even in $h$, while $\z'(\b,h)$ is odd, we get:
\begin{eqnarray}
\ch^{int}(\b,0)
&=&\b^{-1}\dpr^2_h\log{\rm Tr}\,\sum_{\t=\pm1}
\Big(e^{-\b H_0+\b\l\s^z\t+\b h(\s^z+\t)}\Big)\Big|_{h=0}\nn\\
&\defi&\b^{-1}\dpr^2_h\log Z^{int}(\b,h,\l)\Big|_{h=0}
\Eq{eG.7}\\
&=&\b^{-1}\Big[
\sum_\t\frac{\z''+\z' \b \t +(\z'+\b\t\z)\b\t}{Z^{int}}
-\Big(\sum_\t\frac{(\z'+\b\t\z)}{Z^{int}}\Big)^2\Big]_{h=0}
\nn\\
&=&\ch(\b,|\l|)+\b(m(\b,|\l|)+1)^2\tende{\b\to\infty}+\infty\nn\end{eqnarray}
Since $\ch^{int}(\b,0)$ is even in $\l$, it diverges for $\b\to\infty$
independently of the sign of $\l$, while $\ch(\b,0)$ is finite. Hence, the
model yields Pauli's paramagnetism, without a Kondo effect.  \*

{\0{\it Remarks:} (1) Finally an analysis essentially identical to the above
can be performed to study the model in Eq.\equ{e2.1} {\it without impurity}
(and with or without spin) to check that the magnetic susceptibility to a
field $h$ acting only at a single site is finite: the result is the same as
that of the XY model above: the single site susceptibility is finite and,
up to a factor $2$, given by the same formula $\ch(\b,0)=\frac{4\sinh
  \b}{1+\cosh\b}$.}  \*

{\0(2) The latter result makes clear both the essential roles for the Kondo
effect of the spin and of the noncommutativity of the impurity spin
components.}

%%%%%%%%%%%%%%%%%%%%%%%%%%%%%%%%%%%%%%
%%%%%%%%%%%%%%%%%%%%%%%%%%%%%%%%%%%%%
\section{Some details on App.\ref{appG}}
\label{appH}
%%%%%%%%%%%%%%%%%%%%%%%%%%%%%%%%%%%%%%
%%%%%%%%%%%%%%%%%%%%%%%%%%%%%%%%%%%%%%

The definition of $H_h$ has to be supplemented by a boundary condition to
give a meaning to $\Bs_{L+1}$. If $\s^\pm_n=(\s^x\pm i\s^y_n)/2$ define
$\NN_{<n}$ as $\sum_{i<n}\s^+_i\s^-_i=\sum_{i<n}\NN_i$ and $\NN=\NN_{\le
  L}$. Then set as boundary condition

\begin{equation}
\s_{L+1}^\pm\defi -(-1)^\NN\s^\pm_{1\Eq{eH.1}}
\end{equation}
(parity-antiperiodic b.c.) so that $H_h$ becomes
\begin{samepage}\predisplaypenalty0\postdisplaypenalty0
\begin{eqnarray}
H_h&=&-h (2\s^+_\n\s^-_\n - 1)
- \frac12 \sum_{n=1}^{L-1} (\s^+_n(-1)^{\NN_n}\s^-_{n+1}+
\s^-_n(-1)^{\NN_n}\s^+_{n+1})
\Eq{eH.2}\\&&
-\frac12 (\,\s^+_L(-1)^{\NN_L}(-\s^-_{1})
+\s^-_L(-1)^{\NN_L}(-\s^+_{1})\,).
\nn\end{eqnarray}
\end{samepage}
Introducing the Pauli-Jordan transformation
\be a^\pm_n=(-1)^{\NN_{<n}}\s^\pm_n,\qquad a^\pm_{L+1}=-a^\pm_1.
\Eq{eH.3}\ee
In these variables
\be H_h={-h(2 a^+_\n a^-_\n-1) -\frac12} \sum_{n=1}^{L-1} (a^+_na^-_{n+1}-
a^-_n a^+_{n+1})\Eq{eH.4}\ee
Assume $L=$even and let $I\defi\{q| q= \pm\frac{(2n+1)\p}L, \,
n=0,1,\ldots,\frac{L}2 -1\}$; then
\begin{eqnarray}
H_h&=&\sum_q {(-\cos q)} \,A^+_qA^-_q
-\frac{h}L\sum_{q,q'} (2A^+_qA^-_{q'} e^{i(q-q')\nu}
-1)\Eq{eH.5}\\
A^\pm_q&\defi&\frac1{\sqrt L}\sum_{n=1}^L e^{\pm i n q} a^\pm _n,\qquad e^{iL
  q}=-1, \quad q\in I\nn\end{eqnarray}
In diagonal form let $U_{jq}$ be a suitable unitary matrix such that
\be H_h=\sum_j\l_j\a^+_j\a^-_j,\qquad{\rm if}\ \a^+_j=
\sum_q U_{jq} A^+_q\Eq{eH.6}\ee
Then $\l_j$ must satisfy
\begin{eqnarray}
&&\Big(- \sum_q \cos q A^+_q A^-_q
-\frac{2h}L\sum_{q,q'}
A^+_q A^-_{q'}e^{i(q-q')\nu}\Big)
\sum_{q''} U_{jq''} A^+_{q''}\ket0
=\l_j \sum_{q''} U_{jq''} A^+_{q''}\ket0
\Eq{eH.7}\\
{\rm hence}\qquad
&&(\l_j+\cos q)U_{jq}e^{-iq\nu}=
-\frac{2h}L\sum_{q''} e^{-iq''\nu} U_{jq''},
\nn\end{eqnarray}
$\forall q\in I$, where we used the fact that $A^-_p A^+_q \ket0
=\d_{p,q}\ket0$. We consider the two cases $\l_j\ne -\cos q$ for all $q\in I$ or
$\l_j=-\cos q_0$ for some $q_0\in I$.

In the first case:
\be
U_{jq}=\frac{e^{iq\nu}}{N(\l_j)}\frac1{\l_j+\cos q}, \quad
{\rm provided}\quad
F_L(\l_j)\defi
1+\frac{2h}L\sum_q\frac1{\l_j+\cos q}=0,
\Eq{eH.8}\ee
where $N(\lambda_j)$ is set in such a way that $U$ is unitary, or,
in the second case,
\be
\l_j=-\cos q_0,\quad
U_{jq}=\frac{e^{iq\nu}}{\sqrt2}(\d_{q,q_0}-\d_{q,-q_0}),\
\ {\rm so\ that}
\sum_{q''}
e^{-iq''\nu}U_{jq''}=0.
\Eq{eH.9}\ee
Since $-\cos q$ takes $\frac12L$ values and the equation $F_L(\l)=0$ has
$\frac L2$ solutions, the spectrum of $H_h$ is completely determined and
given by the $2^L$ eigenvalues
\be \l({\bf n})=\sum_j n_j \l_j,\qquad \V n=(n_1,\ldots,n_L),
\ n_j=0,1\Eq{eH.10}\ee
and the partition function is
\be
\log Z_L(\b,h)=\sum_{q>0} \log(1+ e^{\b \cos q})
+\sum_{j} \log(1+{ e^{-\b \l_j}} )
= \frac12\log Z^0_L(\b)+\sum_{j\in I}
\log (1+{ e^{-\b \l_j}}).
\Eq{eH.11}\ee
On the other hand, since the function $F'_L(z)/F_L(z)$ has $L/2$ poles
  with residue $+1$ (those corresponding to the zeros of $F_L(z)$) and
  $L/2$ poles with residue $-1$ (those corresponding to the poles of
  $F_L(z)$), the contour integral in the r.h.s. of Eq.\equ{eG.3} is equal
  to
\be
\sum_j \log(1+e^{-\b\l_j}) - \sum_{q>0} \log(1+e^{\b\cos q})
=\sum_j \log(1+e^{-\b\l_j}) - \frac12\log Z^0_L(\b)=
\log Z_L(\b,h) -\log Z^0_L(\b).
\Eq{eH.12}\ee

%%%%%%%%%%%%%%%%%%%%%%%%%%%%%%%%%%%%%%
%%%%%%%%%%%%%%%%%%%%%%%%%%%%%%%%%%%%%
\section{{\tt meankondo}: a computer program to compute flow equations}
\label{appI}
%%%%%%%%%%%%%%%%%%%%%%%%%%%%%%%%%%%%%%
%%%%%%%%%%%%%%%%%%%%%%%%%%%%%%%%%%%%%%

The computation of the flow equation Eq.\equ{eC.1} is quite long, but
elementary, which makes it ideally suited for a computer. We therefore
attach a program, called {\tt meankondo} and written by I.Jauslin, used to
carry it out (the computation has been checked independently by the other
authors). One interesting feature of {\tt meankondo} is that it has been
designed in a {\it model-agnostic} way, that is, unlike its name might
indicate, it is not specific to the Kondo model and can be used to compute
and manipulate flow equations for a wide variety of fermionic hierarchical
models. It may therefore be useful to anyone studying such models, so we
have thoroughly documented its features and released the source code under
an Apache~2.0 license. See \url{http://ian.jauslin.org/software/meankondo}
for details.

\begin{acknowledgements}
We are grateful to V. Mastropietro for suggesting the problem and to
 A. Giuliani, V. Mastropietro and R. Greenblatt for
continued discussions and suggestions, as well as to J. Lebowitz for
hospitality and support.
\end{acknowledgements}

%%%%%%%%%%%%%%%%%%%%%%%%%%%%%%%%%%%%%%%%%%%%%%%%%%%%%%%%%%
%%%%%%%%%%%%%%%%%%%%%%%%%%%%%%%%%%%%%%%%%%%%%%%%%%%%%%%%%%
%%%%%
%\bibliographystyle{apsrev}
\bibliographystyle{spmpsci}
\bibliography{0Bib}

\end{document}